\documentclass[11pt,letterpaper,preprintnumbers,nofootinbib,superscriptaddress,aps,prd]{revtex4}

\preprint{TUM-EFT 186/23, FERMILAB-PUB-24-0029-V}

\pdfoutput=1

\usepackage[T1]{fontenc}
\usepackage[utf8]{inputenc}
\usepackage{amsmath,amssymb,amsfonts,mathtools,bm}
\usepackage[hidelinks]{hyperref}
\usepackage[final]{graphicx}
\usepackage{xcolor}
\usepackage{booktabs}
\usepackage{relsize}
\usepackage{lineno}
\usepackage{comment}
\usepackage{braket}

\def\Babar{{\mbox{\slshape B\kern-0.1em{\smaller A}\kern-0.1em B\kern-0.1em{\smaller A\kern-0.2em R}}}}

\usepackage{cancel}

\usepackage{soul}

\usepackage{siunitx} 
\usepackage[caption=false]{subfig}

\setstcolor{red}

\begin{document}

\title{Bottomonium suppression from the three-loop QCD potential}

\author{Nora Brambilla}
\email{nora.brambilla@tum.de}
\affiliation{Technical University of Munich, TUM School of Natural Sciences, Physics Department, James-Franck-Strasse 1, 85748 Garching, Germany}
\affiliation{Institute for Advanced Study, Technische Universität München, Lichtenbergstrasse 2 a, 85748 Garching, Germany}
\affiliation{Munich Data Science Institute, Technische Universität München, Walther-von-Dyck-Strasse 10, 85748 Garching, Germany}

\author{Tom Magorsch}
\email{tom.magorsch@tum.de}
\affiliation{Technical University of Munich, TUM School of Natural Sciences, Physics Department, James-Franck-Strasse 1, 85748 Garching, Germany}

\author{Michael Strickland}
\email{mstrick6@kent.edu}
\affiliation{Department of Physics, Kent State University, Kent, Ohio 44242, United States}

\author{Antonio Vairo}
\email{antonio.vairo@tum.de}
\affiliation{Technical University of Munich, TUM School of Natural Sciences, Physics Department, James-Franck-Strasse 1, 85748 Garching, Germany}

\author{Peter Vander Griend}
\email{peter.vandergriend@uky.edu}
\affiliation{Department of Physics and Astronomy, University of Kentucky, Lexington, Kentucky 40506, United States}
\affiliation{Theoretical Physics Department, Fermilab, P.O. Box 500, Batavia, Illinois 60510, United States}

\begin{abstract}
We compute the suppression of bottomonium in the quark-gluon plasma using the three-loop QCD static potential.
The potential describes the spin-averaged bottomonium spectrum below threshold with a less than 1\% error.
Within potential nonrelativistic quantum chromodynamics and an open quantum systems framework, 
we compute the evolution of the bottomonium density matrix. 
The values of the quarkonium transport coefficients are obtained from lattice QCD measurements of the bottomonium in-medium width and thermal mass shift; 
we additionally include for the first time a vacuum contribution to the dispersive coefficient $\gamma$.
Using the three-loop potential and the values of the heavy quarkonium transport coefficients, we find that the resulting bottomonium nuclear modification factor is consistent with experimental observations, while at the same time reproducing the lattice measurements of the in-medium width. 
\end{abstract}

\maketitle

\section{Introduction}

The study of the deconfined quark-gluon plasma (QGP) is the main goal of heavy-ion collision experiments.
It provides unique insight into some sectors of the phase diagram of Quantum Chromodynamics (QCD) and a window to conditions that existed in the early universe.
Heavy quarkonium suppression in the QGP is one of the core processes to probe the properties of the QGP, as heavy quarkonium is amenable to a clean experimental reconstruction and a robust theoretical description. 
The idea of heavy quarkonium suppression traces its origin to Matsui and Satz,
who proposed that a screening of the quark-antiquark potential for $J/\psi$ mesons traveling through the QGP produced in heavy-ion collisions should lead to a lower number of particles measured compared to proton-proton collisions~\cite{Matsui:1986dk}. 
In the past, this effect has been studied extensively for bottomonium~\cite{Andronic:2024oxz},
where due to the heavy bottom quark mass, various calculational simplifications are possible,
including the use of effective field theories (EFTs)~\cite{Brambilla:2004jw} and open quantum systems~\cite{Brambilla:2016wgg,Brambilla:2017zei,Brambilla:2020qwo,Akamatsu:2020ypb,Yao:2021lus,Miura:2022arv,Brambilla:2022ynh,Brambilla:2023hkw}.
Ever more precise measurements of the heavy quarkonium nuclear suppression factor,
the observable of interest in quantifying heavy quarkonium suppression, by the ALICE~\cite{ALICE:2020wwx}, ATLAS~\cite{ATLAS:2022exb}, and CMS~\cite{CMS:2018zza,CMS:2023lfu} collaborations necessitate matchingly precise theoretical calculations.

To describe the evolution of heavy quarkonium in medium, it is advantageous to use EFTs that exploit the hierarchy of energy scales exhibited by nonrelativistic bound states. 
Integrating out modes associated with the heavy quark mass $M$ from QCD leads to nonrelativistic QCD (NRQCD)~\cite{Caswell:1985ui,Bodwin:1994jh}.
Further, integrating out soft modes associated with the momentum transfer scale given by $M$ times the relative velocity $v$, 
which is also the scale of the inverse of the heavy quark-antiquark distance in the bound state, 
leads to potential NRQCD (pNRQCD)~\cite{Pineda:1997bj,Brambilla:1999xf} and, 
at the Lagrangian level, to the emergence of quarkonium potentials.
For weakly coupled quarkonia, the potentials may be computed in perturbation theory;
at leading order in the nonrelativistic and coupling expansion, 
these are an attractive color-singlet and repulsive color-octet static potential.
Combining nonrelativistic EFTs with the open quantum system (OQS) framework, it is possible to derive a master equation for the evolution of the quarkonium density matrix \cite{Brambilla:2016wgg,Brambilla:2017zei,Brambilla:2022ynh}. 

A weak coupling treatment of the quarkonium soft modes is justified for quarkonia, whose typical radius 
is much smaller than the inverse of the hadronic scale $\Lambda_{\textrm{QCD}}$. 
This is certainly the case for the lowest-lying bottomonium states and may be extended, although more marginally, to all bottomonium states below the open flavor threshold.
At leading order, the weak-coupling static potential coincides with the Coulomb potential.
It is well known that the Coulomb potential describes poorly the quarkonium spectrum~\cite{Brambilla:1999ja}.
However, it has been shown that including higher-order corrections to the static potential,
which is known up to three loops~\cite{Brambilla:1999qa,Anzai:2009tm,Smirnov:2009fh}, 
and subtracting from the perturbative series the leading renormalon contribution by reabsorbing it into a renormalon subtracted mass~\cite{Pineda:2001zq}
leads to a well-behaved perturbative series and a fair reproduction of the bottomonium 
spectrum below threshold (see, for instance, Refs.~\cite{Brambilla:2001fw,Sumino:2016sxe,Peset:2018jkf}).
In this work, we use the three-loop renormalon subtracted version of the QCD static potential 
put forward in Refs.~\cite{Kiyo:2010jm,Pineda:2013lta,Peset:2018jkf,Segovia:2018qzb}.
We treat this potential, together with the kinetic energy, as the leading order contribution 
to the pNRQCD Hamiltonian and, therefore, to the Schr\"odinger equation describing the bound state. For an early application based on the same idea, see Ref.~\cite{Titard:1993nn}.
We further supplement the perturbative static potential with its leading nonperturbative 
correction. This correction has been derived rigorously in pNRQCD and has the form of a harmonic potential, whose strength is given by the time integral of a suitable nonlocal chromoelectric correlator~\cite{Brambilla:1999xf,Brambilla:2000am}.

The Lindblad equation describing the time evolution of the bottomonium 
density matrix in the QGP formed in heavy-ion collisions 
has been derived in pNRQCD in Refs.~\cite{Brambilla:2016wgg,Brambilla:2017zei}.
The Lindblad equation depends on the static potential defined above and on heavy quarkonium transport coefficients. 
The transport coefficients may be extracted from fitting available lattice data of the bottomonium thermal mass shifts and decay widths~\cite{Larsen:2019bwy} with the corresponding quantities computed in pNRQCD.
Finally, we solve the Lindblad equation in the computational setup of Ref.~\cite{Omar:2021kra}.
We work in the strict quantum Brownian limit, i.e., we assume the system correlation time to be much larger than the environment correlation time, which is proportional to the inverse of the temperature, and we neglect higher-order corrections.
We find an accurate description of the LHC data on bottomonium suppression.

The remainder of the paper is structured as follows: Section~\ref{sec:oqs} briefly presents the theoretical background for describing heavy quarkonium suppression using EFTs and open quantum systems.
In section~\ref{sec:pot}, we introduce the three-loop renormalon subtracted static potential and its leading nonperturbative correction,
and in section~\ref{sec:transp}, we discuss the quarkonium transport coefficients used in the simulation. We present our results for the nuclear modification factor in section~\ref{sec:results} and conclude in section~\ref{sec:conc}.

\section{Theory}
\label{sec:theory}

\subsection{Open quantum system formulation of quarkonium suppression}
\label{sec:oqs}

Heavy quarkonium exhibits distinct hierarchically ordered energy scales, making it amenable to an EFT description. 
Using pNRQCD, in Refs.~\cite{Brambilla:2016wgg,Brambilla:2017zei}, a master equation was derived for the density matrix of a quarkonium state traveling through a strongly coupled, thermal QGP 
and satisfying the conditions 
\begin{equation}
    M \gg 1/a_0 \gg \pi T \sim g T , \Lambda_{\textrm{QCD}}\gg E \, ,
\label{eq:scales}
\end{equation}
where $a_0$ is the quarkonium Bohr radius, $T$ the QGP temperature, and $E$ the typical quarkonium energy.
Because $1/E$ is the system correlation time, the strict Brownian limit corresponds to expanding in $E/(\pi T)$ and keeping only the leading term;
beyond the strict Brownian limit, the Lindblad equation has been studied in Ref.~\cite{Brambilla:2022ynh} at order $E/(\pi T)$. 
In the strict Brownian limit, the master equation can be cast into a Lindblad form~\cite{Lindblad:1975ef,Gorini:1975nb}
\begin{equation}
    \frac{d\rho(t)}{dt}=-i[H,\rho(t)]+\sum_n\left[C^n_i\rho(t)C^{n\dagger}_i-\frac{1}{2}\left\{C^{n\dagger}_iC^n_i,\rho(t)\right\}\right],
    \label{eq:lindblad}
\end{equation}
with Hamiltonian
\begin{equation}
H=
\begin{pmatrix} 
h_s+ \displaystyle\frac{r^2}{2}\gamma & 0 \\
0 & \displaystyle h_o+\frac{N^2_c-2}{2(N^2_c-1)}\frac{r^2}{2}\gamma
\end{pmatrix},
\label{eq:Hamiltonian}
\end{equation}
and Lindblad operators
\begin{align}
    C^0_i &= \sqrt{\frac{\kappa}{N^2_c-1}}r_i\begin{pmatrix}
        0 & 1\\
        \sqrt{N^2_c-1} & 0
    \end{pmatrix},\\
    C^1_i &= \sqrt{\frac{\kappa(N^2_c - 4)}{2(N^2_c-1)}}r_i\begin{pmatrix}
        0 & 0\\
        0 & 1
    \end{pmatrix}.\label{eq:c2}
\end{align}
The operators $h_s$ and $h_o$ encode the contributions from the energy modes that have been integrated out in constructing pNRQCD.
These are, according to the hierarchy of Eq.~\eqref{eq:scales},  all the modes associated with energy scales larger than the thermal scales and $\Lambda_{\textrm{QCD}}$.
By construction, therefore, $h_s$ and $h_o$ are the in-vacuum, perturbative Hamiltonians of a heavy quark-antiquark pair in a color singlet and octet configuration, respectively,
\begin{equation}
h_s = \frac{\bm{p}^2}{M} + V_s^{\textrm{pert}}(r), \qquad\qquad  h_o = \frac{\bm{p}^2}{M} + V_o^{\textrm{pert}}(r), 
\end{equation}
where $\bm{r}$ is the heavy quark-antiquark distance, $\bm{p} = -i \bm{\nabla}_r$ the relative momentum, 
$V_s^{\textrm{pert}}$ the color singlet and $V_o^{\textrm{pert}}$ the color octet static potential computed in perturbative QCD.
The number of colors is $N_c=3$, and, for further use, the Casimir of the adjoint representation is $C_A=3$, the Casimir of the fundamental representation is $C_F=4/3$ and the color matrix normalization is $T_F=1/2$.

The low-energy and thermal dynamics in the evolution equation are encoded in transport coefficients.
In deriving Eqs.~\eqref{eq:Hamiltonian}-\eqref{eq:c2}, we have considered only the heavy quarkonium momentum diffusion coefficient $\kappa$ and its dispersive counterpart, $\gamma$.
The transport coefficients $\kappa$ and $\gamma$ are defined in terms of chromoelectric correlators as 
\begin{align}
  \kappa&=\frac{g^2}{3N_c}\text{Re}\int^\infty_0 dt \, \braket{E^{a\,i}(t)\Omega(t,0)^{ab}E^{b\,i}(0)}, \label{eq:kappa} \\
  \gamma&=\frac{g^2}{3N_c}\text{Im}\int^\infty_0 dt \, \braket{E^{a\,i}(t)\Omega(t,0)^{ab}E^{b\,i}(0)}, \label{eq:gamma}
\end{align}
where $\displaystyle \Omega(t_2,t_1)^{ab}=P\exp\left(-ig\int^{t_2}_{t_1}dt A^c_0(t)(-if^{abc})\right)$ is an adjoint Wilson line.

The Lindblad equation provides the full three-dimensional evolution of the quarkonium density matrix. 
Utilizing the spherical symmetry of the system, we follow the procedure given in detail in Appendix C of Ref.~\cite{Brambilla:2022ynh} to bring the evolution equation into the form of a one-dimensional Lindblad equation describing the evolution of the radial wave function; this greatly reduces the computational cost of solving the Lindblad equation. 
The solution of the resulting equation is found by using the open-source code \texttt{QTraj}~\cite{Omar:2021kra},
which employs the Monte Carlo wave function method~\cite{Molmer:1993ltv} to solve the Lindblad equation in a computationally efficient and embarrassingly parallel manner.

\subsection{Heavy quarkonium potential}
\label{sec:pot}

In previous works~\cite{Brambilla:2020qwo,Brambilla:2022ynh,Brambilla:2023hkw},
the perturbative color-singlet and color-octet potentials were taken at leading order in $\alpha_s$, i.e., $V_s^{\textrm{pert}}(r) \approx V_s^c(r) \equiv -4\,\alpha_s/(3r)$ and $V_o^{\textrm{pert}}(r) \approx V_o^c(r) \equiv \alpha_s/(6r)$,
with $\alpha_s$, the strong coupling, evaluated at the inverse of the Bohr radius $a_0$ determined from the solution of the self-consistency equation
\begin{equation}
    a_0 = \frac{3}{2\,M_b\,\alpha_s(1/a_0)} \, .
    \label{eq:self-cons}
\end{equation}
Taking $M_b=\SI{4.850}{\giga\eV}$ in order to reproduce the $\Upsilon(1S)$ mass from the solution of the Schr\"odinger equation with $V_s^{\textrm{pert}}\approx V_s^c$  and solving 
Eq.~\eqref{eq:self-cons} by running $\alpha_s$ at one-loop accuracy using RunDec~\cite{Herren:2017osy} from $\alpha_s(M_Z=\SI{91.1876+-0.0021}{\giga\eV})=\num{0.1181+-0.0011}$, we find $1/a_0=\SI{1.11}{\giga\eV}$.

To improve on this description, we consider here the case where $V_s^{\textrm{pert}}$ and $V_o^{\textrm{pert}}$ incorporate the three-loop expression.
The three-loop color singlet potential has been computed in Refs.~\cite{Anzai:2009tm,Smirnov:2009fh} and the three-loop color octet potential in Ref.~\cite{Anzai:2013tja}.
At three loops (3L), both the singlet and octet potential are infrared divergent with the divergence reabsorbed into nonperturbative corrections to the energy~\cite{Brambilla:1999qa,Brambilla:1999xf};
we call $\nu_{\textrm{us}}$ the corresponding renormalization scale.
Higher-loop corrections to the static potential are poorly convergent, 
a behavior that may be traced back to the order $\Lambda_{\textrm{QCD}}$ renormalon affecting the potential.
The renormalon may be subtracted and reabsorbed into a redefinition of the heavy quark mass~\cite{Pineda:1998PhD,Hoang:1998nz}.
We adopt the renormalon subtraction scheme RS$^\prime$ introduced in Ref.~\cite{Pineda:2001zq};
we call $\nu_f$ the renormalon factorization  scale.
At very short distances, potentially large logarithms of the form $\log(\nu r)$, where $\nu$ is the renormalization scale of the strong coupling, are resummed in the running of $\alpha_s$ by setting $\nu=1/r$.
At larger distances, keeping the renormalization scale constant ensures the convergence of the perturbative expansion.
We call $\nu_r$ the scale separating the two regions, while still keeping both in the perturbative regime.
The resulting potential has then the form~\cite{Kiyo:2010jm,Pineda:2013lta,Peset:2018jkf,Segovia:2018qzb}
\begin{equation}
  V_s^{\textrm{pert}}(\nu,\nu_f,\nu_r,\nu_{\textrm{us}},r) \approx V_s^{\textrm{3L}}(\nu,\nu_f,\nu_r,\nu_{\textrm{us}},r)
  = \begin{cases}
    \sum^3_{k=0}V^{(k)}_{s,\text{RS}^\prime}\alpha^{k+1}_s(1/r) \quad \text{if}\;\, r< \nu^{-1}_r \\
    \sum^3_{k=0}V^{(k)}_{s,\text{RS}^\prime}\alpha^{k+1}_s(\nu) \qquad \text{if}\; r> \nu^{-1}_r \end{cases}.
\label{eq:singlet-pert}
\end{equation}
The explicit expression of the coefficients $V^{(k)}_{s,\text{RS}^\prime}$ can be found in Appendix~\ref{app:pot}.
The potential depends on the four scales $\nu$, $\nu_f$, $\nu_r$, and $\nu_{\textrm{us}}$.
We set $\nu_f=\nu_{\textrm{us}}=\SI{1}{\giga\eV}$ and $\nu_r = \nu =1/a_{0}=\SI{1.12}{\giga\eV}$, where $1/a_{0}$ is calculated using Eq.~\eqref{eq:self-cons} for $M_b=\SI{4.921}{\giga\eV}$ (for the choice of $M_b$, see the discussion at the end of this subsection). 
We evaluate $\alpha_s$ using RunDec~\cite{Herren:2017osy} from $\alpha_s(M_Z)$ and evaluate $\alpha^{k+1}_s$  to $(4-k)$-loop accuracy in Eq.~\eqref{eq:singlet-pert}.

In the scale setting of Eq.~\eqref{eq:scales}, the quarkonium potential also gets nonperturbative and thermal corrections.
The leading one is encoded in the term proportional to $\gamma r^2 /2$ in the Hamiltonian $H$ given in Eq.~\eqref{eq:Hamiltonian}.\footnote{
Higher order terms not considered here ensure the cancellation of the $\nu_{\textrm{us}}$ dependence of the perturbative 
potential~\cite{Brambilla:1999xf}.}
The in-vacuum, nonperturbative (np) correction to the color singlet static potential, reads~\cite{Brambilla:1999xf,Brambilla:2000am}
\begin{equation}
V^{\text{np}}_s(r) = \frac{r^2}{2}\,\gamma(T=0) \,.
\label{eq:singlet_non-pert}
\end{equation}
The quantity $\gamma(T=0)$ stands for the in-vacuum, zero-temperature part 
of the chromoelectric correlator defined in Eq.~\eqref{eq:gamma}.
The fact that $\gamma(T=0)$ is different from zero should be contrasted with the fact that  
the quarkonium momentum diffusion coefficient $\kappa$ vanishes in vacuum 
as a consequence of quarkonium decay into an unbound heavy quark-antiquark pair being kinematically forbidden in vacuum.
An estimate of $\gamma(T=0)$ is provided in Eq.~\eqref{gammavacuum} of section~\ref{sec:transp}.
The full in vacuum color-singlet potential that we consider here is then 
\begin{equation}
    V^{\text{3L+np}}_s(r) = V^{\text{3L}}_s(\nu,\nu_f,\nu_r,\nu_{\textrm{us}},r) + V^\text{np}_s(r) \, .
\label{eq:fullpot}
\end{equation}
Its perturbative part is the three-loop, renormalon subtracted, renormalization group improved color-singlet potential defined in Eq.~\eqref{eq:singlet-pert}.
Its nonperturbative part is given in Eq.~\eqref{eq:singlet_non-pert}.

Similarly we define a three-loop, renormalon subtracted, renormalization group improved color-octet potential~\cite{Bali:2003jq,Anzai:2013tja}
\begin{equation}
V_o^{\textrm{pert}}(\nu,\nu_f,\nu_r,\nu_{\textrm{us}},r) \approx V_o^{\textrm{3L}}(\nu,\nu_f,\nu_r,\nu_{\textrm{us}},r)
=\begin{cases}
  \sum^3_{k=0}V^{(k)}_{o,\text{RS}^\prime}\alpha^{k+1}_s(1/r) \quad \text{if}\;\, r< \nu^{-1}_r \\
  \sum^3_{k=0}V^{(k)}_{o,\text{RS}^\prime}\alpha^{k+1}_s(\nu) \qquad \text{if}\; r> \nu^{-1}_r 
\end{cases}.
\label{eq:octet}
\end{equation}
The coefficients $V^{(k)}_{o,\text{RS}^\prime}$ are listed in Appendix~\ref{app:pot}.

\begin{figure}[ht]
    \centering
    \includegraphics[width=0.7\textwidth]{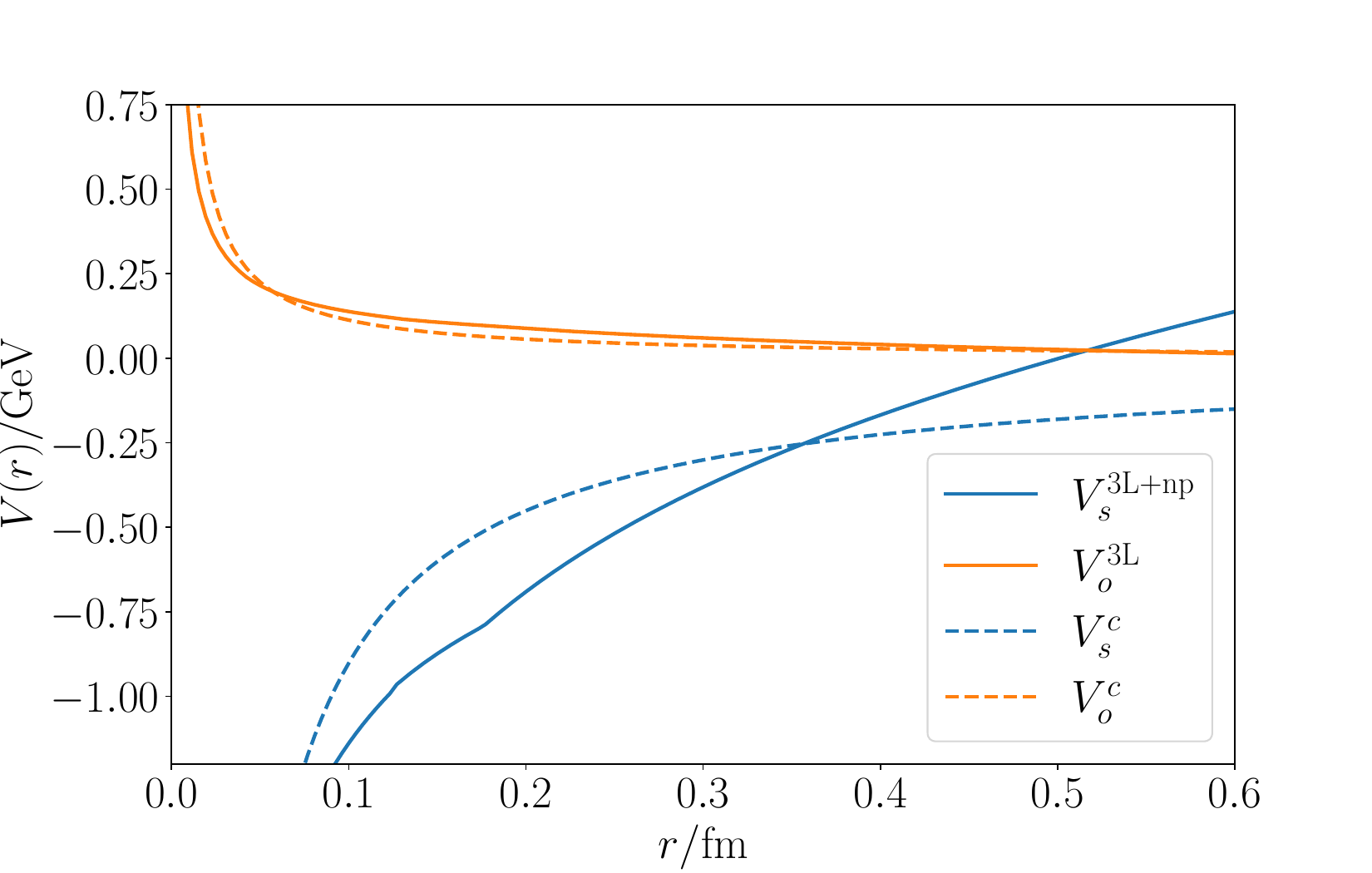}
    \caption{The singlet and octet Coulomb potentials $V^c_s$ and $V^c_o$ (dashed lines) compared to the improved potentials $V^\text{3L+np}_s$ and $V^\text{3L}_o$ (solid lines).
      For the improved potentials, there is a discontinuity in the first derivative at \mbox{$1/\nu_r=\SI{0.197}{\femto\meter}$}.
      This can be seen by the kink in the solid blue curve, 
      while it is barely noticeable in the solid orange curve that describes the octet potential.}
    \label{fig:potplot}
\end{figure}

A comparison between the potentials $V^{\text{3L+np}}_s$, $V^{\text{3L}}_o$ and the Coulomb potentials $V^c_s$, $V^c_o$ is shown in Fig.~\ref{fig:potplot}.
Since the expressions of the potentials in Eqs.~\eqref{eq:fullpot} and~\eqref{eq:octet} rely on assuming the soft modes to be perturbative, the curves are theoretically justified only at distances shorter than $1/ \Lambda_{\textrm{QCD}}$, which we may indicatively take to be about 0.3~fm.
The curves for $V^{\text{3L}}_o$ and $V^c_o$ exhibit no qualitative difference, while $V^{\text{3L+np}}_s$ shows a much steeper slope than $V^c_s$ at distances between \mbox{0.15~fm} and 0.3 fm (and larger).
The large, positive slope of the potential $V^{\text{3L+np}}_s$ at those distances is suggestive of an onset of confinement.

The bottomonium spectrum follows from solving the radial Schr\"odinger equation
\begin{equation}
-\frac{1}{M_b}\frac{d^2u_{nl}(r)}{dr^2} + \left(V_s(r)+\frac{l(l+1)}{M_br^2}u_{nl}(r)\right)=E_{nl}u_{nl}(r) \, ,
\label{eq:schroedinger}
\end{equation}
where $u_{nl}(r)$ is the reduced radial wave function of the state, $E_{nl}$ the corresponding binding energy (hence the mass of the state is $2M_b+E_{nl}$), and $V_s$ the color singlet static potential.
The $b$-mass $M_b$ is fixed by fitting the $1S$ mass to the PDG value $M(1S)=\SI{9.445}{\giga\eV}$.
The bottomonium masses are understood as spin averaged.
The $b$-mass that follows from solving the Schr\"odinger equation for $V_s = V_s^c$ is $M_b=\SI{4.850}{\giga\eV}$, 
whereas the one that follows from solving the Schr\"odinger equation for $V_s = V^\text{3L+np}_s$, 
with $\gamma(T=0)$ given in Eq.~\eqref{gammavacuum}, is $M_b=\SI{4.921}{\giga\eV}$.
The results for the spectrum are shown in Table~\ref{tab:spectrum}.
Since we neglect relativistic corrections to the potential, which are $1/M_b$ suppressed, 
we do not expect to reproduce the spectrum exactly;
nevertheless, we note that in contrast to the Coulomb potential, the potential $V^\text{3L+np}_s$ yields a spectrum accurate to within 1\%.
Moreover, the potential $V^\text{3L}_s$ has been instrumental in reproducing the fine structure of  the bottomonium spectrum~\cite{Peset:2018jkf}, electromagnetic decays~\cite{Kiyo:2010jm} and electromagnetic transition widths~\cite{Pineda:2013lta,Segovia:2018qzb}.

\begin{table}[t]
    \centering
    \begin{tabular}{crrr}
         & PDG & ~$V^{c}_s$ & $V^{\text{3L+np}}_s$ \\ \hline
        $M(1S)/\,\si{\giga\eV}$ & ~$\num{9.445}$ & 
        ~$\num{9.445}$ & ~$\num{9.445}$\\ \hline
        $M(2S)/\,\si{\giga\eV}$ & ~$\num{10.017}$ & ~$\num{9.635}$ & ~$\num{10.042}$\\ \hline
        $M(3S)/\,\si{\giga\eV}$ & ~$\num{10.355}$ & ~$\num{9.670}$ & ~$\num{10.395}$\\ \hline
        $M(1P)/\,\si{\giga\eV}$ & ~$\num{9.888}$ 
        & ~$\num{9.635}$ & ~$\num{9.887}$\\ \hline
        $M(2P)/\,\si{\giga\eV}$ & ~$\num{10.251}$ & ~$\num{9.670}$ & ~$\num{10.279}$ \\ 
    \end{tabular}
    \caption{The spin-averaged bottomonium spectrum from PDG data~\cite{ParticleDataGroup:2020ssz}, 
     from solving the Schr\"odinger equation with the Coulomb potential $V^\text{c}_s$,
     and from solving the Schr\"odinger equation with the potential $V^\text{3L+np}_s$ and $\gamma(T=0)$ as in Eq.~\eqref{gammavacuum}.
     The $b$-mass is fixed by fitting the $1S$ prediction to the PDG value yielding $M_b=\SI{4.850}{\giga\eV}$ for $V^\text{c}_{s}$ and $M_b=\SI{4.921}{\giga\eV}$ for $V^\text{3L+np}_s$.}
    \label{tab:spectrum}
\end{table}

\subsection{Transport coefficients}
\label{sec:transp}
According to Eqs.~\eqref{eq:lindblad}-\eqref{eq:c2}, 
the interaction between heavy quark-antiquark pairs and the QGP is encoded 
in the transport coefficients $\kappa$ and $\gamma$, defined in Eq.~\eqref{eq:kappa} and~\eqref{eq:gamma}, respectively~\cite{Brambilla:2016wgg,Brambilla:2017zei,Brambilla:2019tpt,Eller:2019spw}.
In contrast to $\kappa$, $\gamma$ contains vacuum contributions.
To isolate the vacuum part, we write 
\begin{align}
    \kappa &= \hat\kappa T^3 \, ,\\
    \gamma &= \gamma(T=0) + \hat\gamma T^3 \, ,
\end{align}
where $\gamma(T=0)$ denotes the vacuum contibution to $\gamma$ while $\hat\kappa$ and $\hat\gamma$ are dimensionless coefficients of the thermal contributions.
Up to now, there are no direct lattice QCD determinations of the quantities $\kappa$ and $\gamma$.
We note that lattice measurements of a related quantity, the heavy quark momentum diffusion coefficient, have existed for a number of years (for a recent determination see Ref.~\cite{Brambilla:2020siz}) and anticipate lattice measurements of $\kappa$ and $\gamma$ in the near future.

To obtain an estimate for $\gamma(T=0)$, we parametrize the nonperturbative part of the chromoelectric correlator as 
\begin{equation}
\bra{0} E^{a\,i}(t)\Omega(t,0)^{ab}E^{b\,i}(0) \ket{0}=\bra{0} E^2(0) \ket{0}\mathrm{e}^{-i\Lambda_E t} \, ,
\label{eq:condensate}
\end{equation}
where $\ket{0}$ is the vacuum state, $\bra{0} E^2(0) \ket{0}$ the chromoelectric condensate, and $\Lambda_E$ the gluelump mass~\cite{Foster:1998wu,Brambilla:1999xf,Brambilla:2000am}.
This parameterization interpolates in the simplest possible way between the condensate at $t=0$ and a large time behavior where 
the spectral decomposition of the correlator is dominated by the lowest-lying gluelump with the quantum numbers of the chromoelectric field.
The perturbative part of the correlator is scaleless and proportional to $1/t^4$ and, therefore, vanishes in $\gamma(T=0)$ when regularized in dimensional regularization.
We take the gluon condensate to be $\bra{0} \alpha_s G^2 \ket{0}=\SI{6.4+-0.3e-2}{\giga\eV\tothe{4}}$~\cite{Narison:2023npv}, 
which leads to $\bra{0} g^2E^2(0) \ket{0} =\SI{-0.2}{\giga\eV\tothe{4}}$, and  the chromoelectric gluelump mass $\Lambda_E=\SI{1.25}{\giga\eV}$~\cite{Bali:2003jq,Herr:2023xwg}. 
Inserting Eq.~\eqref{eq:condensate} into Eq.~\eqref{eq:gamma}, performing the integral, and regularizing at infinity, we obtain
\begin{equation}
\gamma(T=0)=-\frac{\bra{0} g^2E^2(0) \ket{0}}{3 N_c\Lambda_E}=\SI{0.017}{\giga\eV\tothe{3}} \, .
\label{gammavacuum}
\end{equation}

The thermal part $\hat\gamma$ can be extracted from lattice measurements of the in-medium heavy quarkonium mass shift~\cite{Larsen:2019bwy}. 
In our setting, the thermal mass shift is given by~\cite{Brambilla:2019tpt}
\begin{equation}
\delta M = \frac{\hat\gamma \,T^3}{2} \, \int_0^\infty dr \, u_{nl}^* r^2 \, u_{nl}\,,
\label{eq:thermalmass}
\end{equation}
where $u_{nl}$ is the solution of the Schr\"odinger equation \eqref{eq:schroedinger}.
Since the results reported in Ref.~\cite{Larsen:2019bwy} suggest a vanishing thermal mass shift, we set $\hat\gamma=0$. 
Note that the vacuum part $\gamma(T=0)$ is not accessible through the thermal mass shift.

\begin{figure}[ht]
    \centering
    \includegraphics[width=0.495
\textwidth]{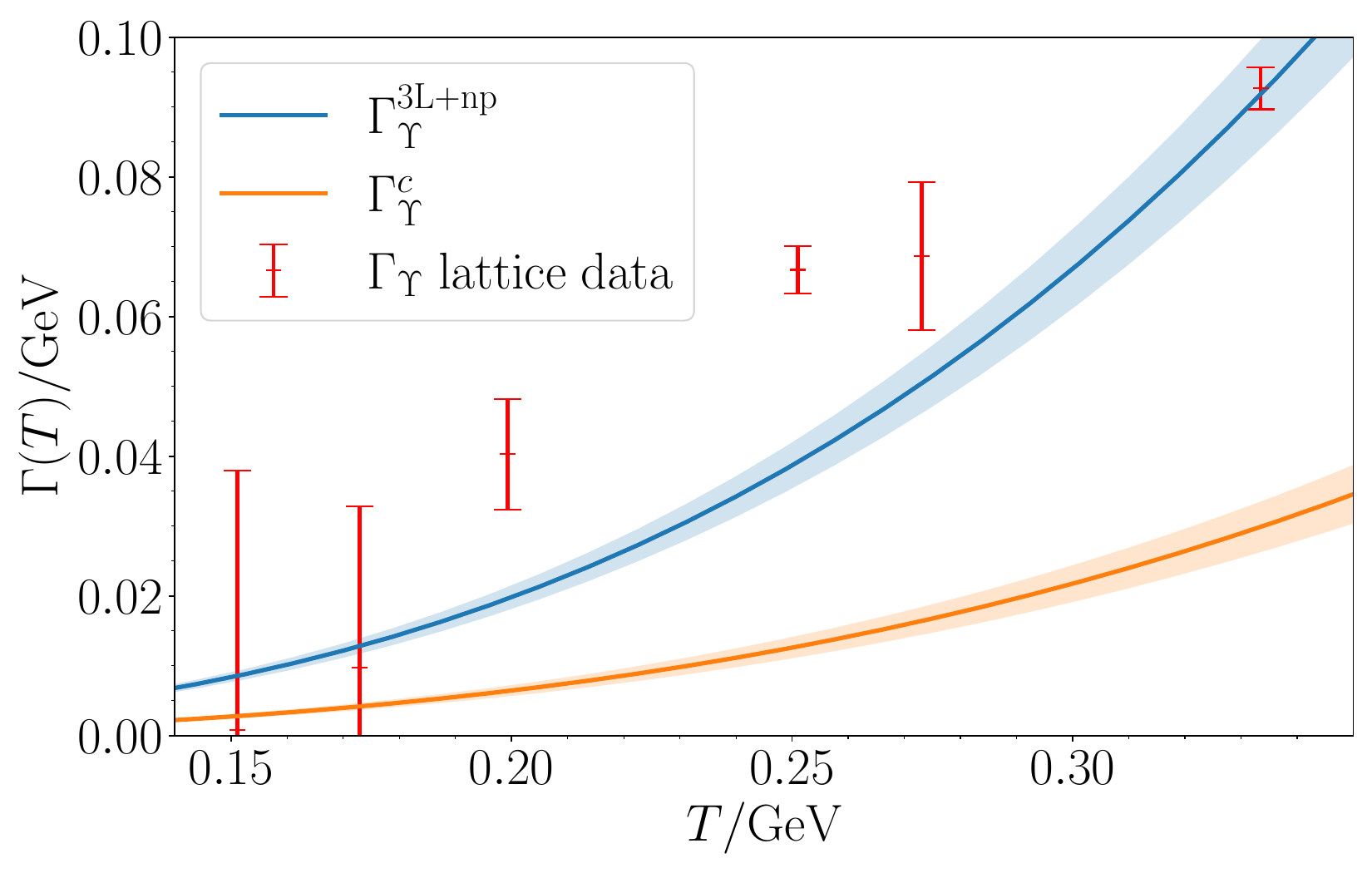}
    \includegraphics[width=0.495\textwidth]{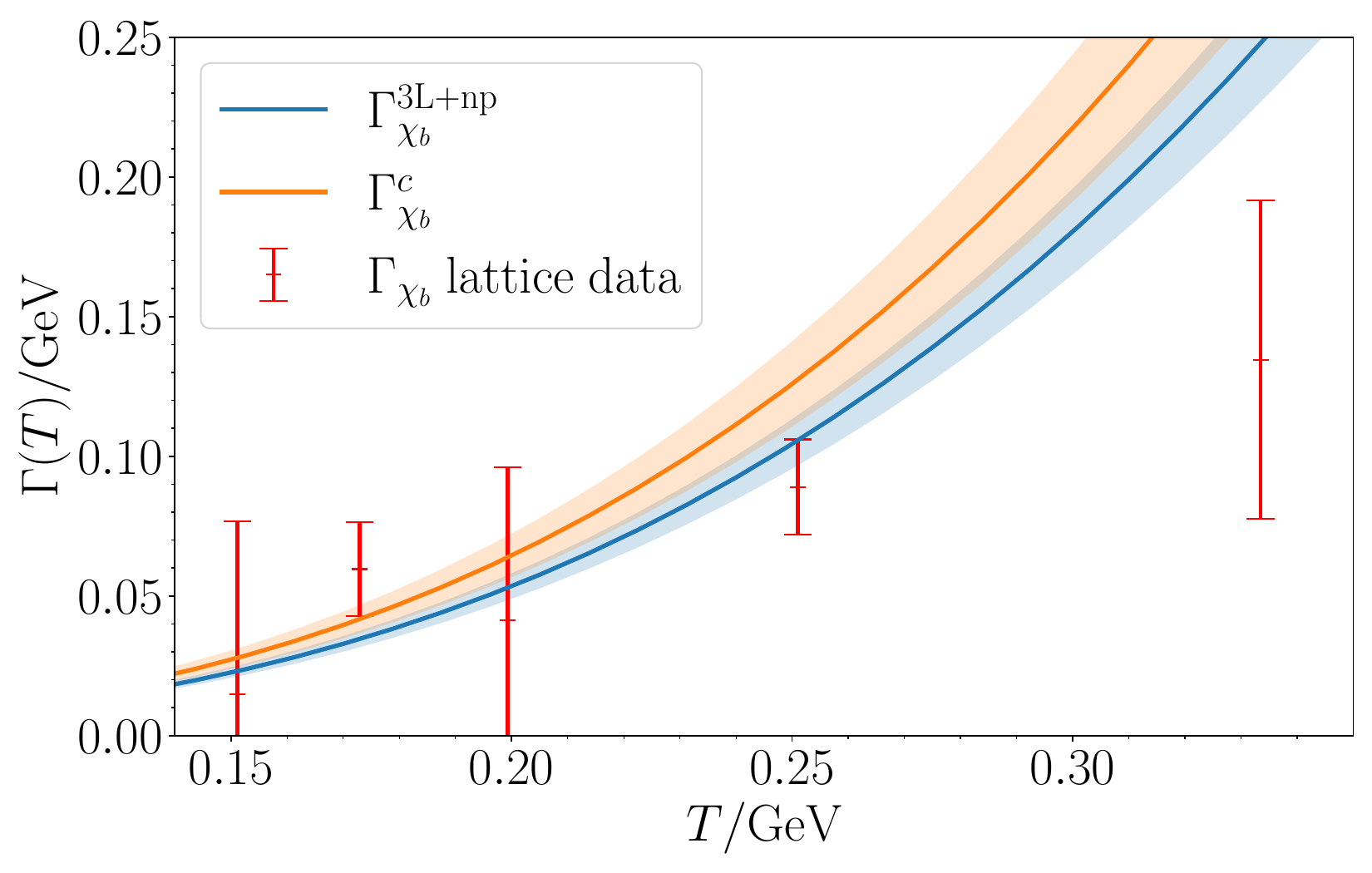}
    \caption{The calculated widths $\Gamma$ for the Coulomb potential $V_s^c$ 
      and the improved potential $V^\text{3L+np}_s$ compared to the lattice data from Ref.~\cite{Larsen:2019bwy}.
      The left panel shows calculations for $l=0$ states compared to $\Upsilon$ data, 
      and the right panel shows calculations for  $l=1$ states compared to $\chi_b$ lattice data.
      In both cases, we use the extracted and averaged values $\hat\kappa=\num{0.33}\pm 0.04$ for the Coulomb potential 
      and $\hat\kappa=\num{1.88} \pm 0.16$ for the potential $V^\text{3L+np}_s$.
      The lattice widths displayed are rescaled by $\left({M^\text{lattice}_b}/{M_b}\right)^2$ with  $M_b=\SI{4.921}{\giga\eV}$.}
    \label{fig:width}
\end{figure}

Likewise, we extract values for $\hat\kappa$ by calculating the width $\Gamma$ and fitting it to the lattice measurements of the in-medium width reported in Ref.~\cite{Larsen:2019bwy}. 
In our setting, the quarkonium width, which is entirely thermal, is given by~\cite{Brambilla:2019tpt}
\begin{equation}
\Gamma = \hat\kappa \,T^3 \, \int_0^\infty dr \, u_{nl}^* r^2 \, u_{nl}\,.
\label{eq:width}
\end{equation}
Since different $b$-masses were used in the lattice calculation~\cite{Larsen:2019bwy},
we rescale the reported lattice widths by a factor $\left({M^\text{lattice}_b}/{M_b}\right)^2$, where $M^\text{lattice}_b$ denotes the $b$-mass used in the lattice computation (see table II of Ref.~\cite{Larsen:2019bwy}),
and $M_b$ is the $b$-mass used in our computation. 
In practice, we rescale using $M_b=\SI{4.921}{\giga\eV}$, since rescaling with the $b$-mass determined for the Coulomb potential, $M_b=\SI{4.850}{\giga\eV}$, leads to compatible widths within the error bars.
Due to the Brownian limit, the width in Eq.~\eqref{eq:width} may not be accurate for the lowest temperatures; we have checked, however, that excluding the lattice data for $T<\SI{200}{\mega\eV}$ in the fit does not change qualitatively our results.
In the case of the potential $V^\text{3L+np}_s$,
a weighted least squares fit of $\hat\kappa$ from the $\Upsilon$ data gives $\hat\kappa_\Upsilon=\num{2.08+-0.20}$, and, from the $\chi_{b1}$ data, $\hat\kappa_\chi=\num{1.52+-0.26}$. 
The weighted average of these two determinations is $\hat\kappa=\num{1.88+-0.16}$. 
Applying the same procedure for the Coulomb potential, we find $\hat\kappa_\Upsilon=\num{1.16+-0.11}$ and $\hat\kappa_\chi=\num{0.23+-0.04}$ leading to the average $\hat\kappa=\num{0.33+-0.04}$. 
We plot the temperature dependence of the width for the lowest-lying states with $l=0$ and $l=1$ in the left and right panels of Fig.~\ref{fig:width}, respectively. 
The plots show with an orange line the width for the final average $\hat\kappa=\num{0.33}\pm 0.04$, obtained from the Coulomb potential, 
and with a blue line the width for the final average $\hat\kappa=\num{1.88\pm 0.16}$, obtained from the improved potential $V^\text{3L+np}_s$. 
We see that the Coulomb fit is unable to describe both the $\Upsilon$ and $\chi_b$ data at the same time, 
while the fit based on the improved potential $V^\text{3L+np}_s$ provides a more reasonable description of the lattice data in both cases.

\section{Results}
\label{sec:results}

\begin{figure}[ht]
    \centering
    \includegraphics[width=0.65\textwidth]{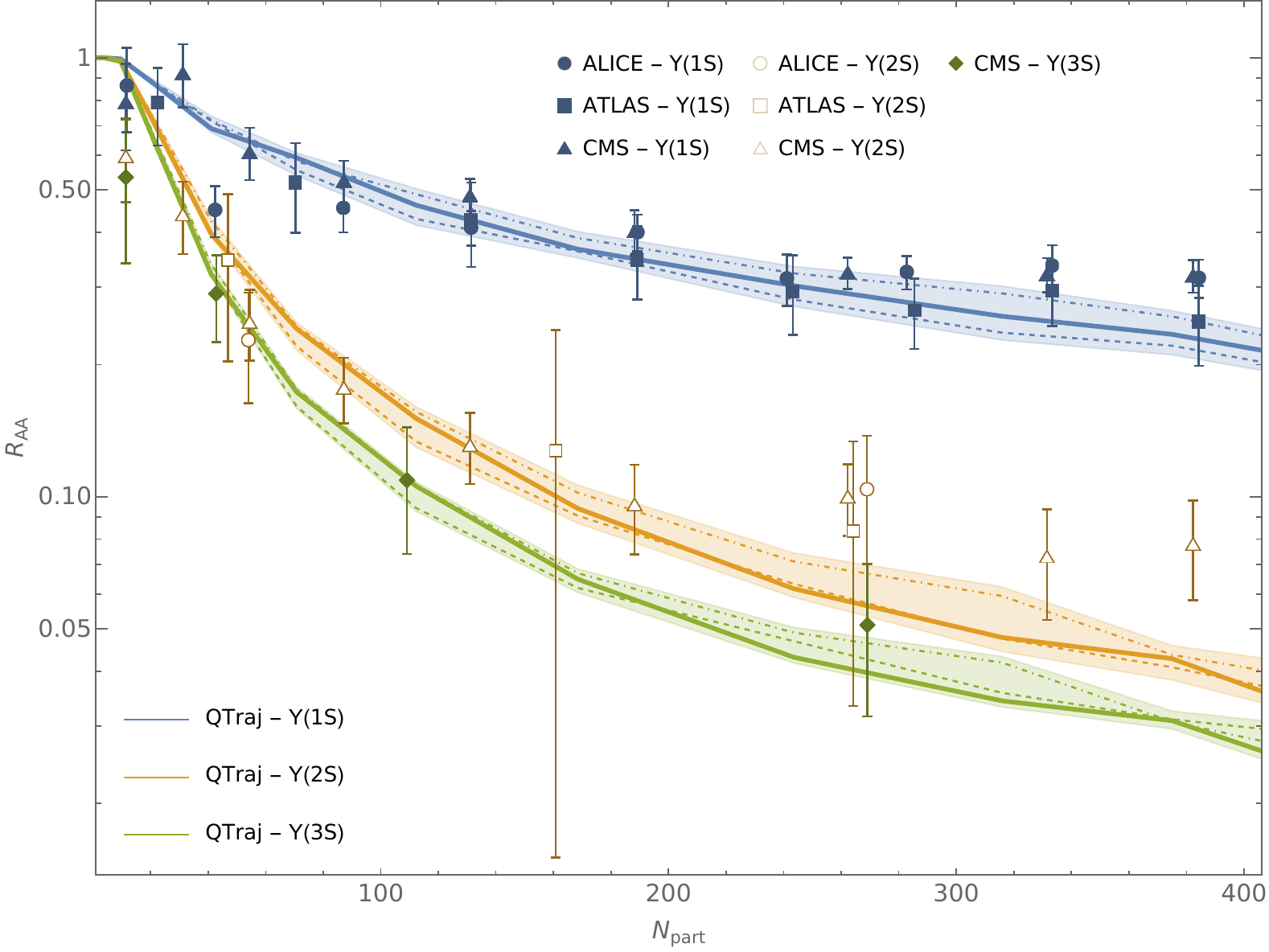}
    \caption{The nuclear modification factor $R_{AA}$ for the $1S$, $2S$, and $3S$ bottomonium states as a function of the number of participating nucleons $N_\text{part}$. 
    The colored solid lines show results obtained with $\hat\kappa=\num{1.88}$, $\gamma(T=0)=\SI{0.017}{\giga\eV\tothe{3}}$ and $\hat\gamma=0$ using the three-loop potentials $V^\text{3L}_\text{s}$ and $V^\text{3L}_\text{o}$ in the evolution equation. 
    The colored bands correspond to the statistical uncertainties and the uncertainties from $\hat\kappa$. 
    The dash-dotted and dashed lines indicate the curves of $\hat\kappa=\num{1.72}$ and $\hat\kappa=\num{2.04}$, respectively. 
    The data are taken from the ALICE~\cite{ALICE:2020wwx}, ATLAS~\cite{ATLAS:2022exb}, and CMS~\cite{CMS:2018zza,CMS:2023lfu} experiments.}
    \label{fig:resultsvpert}
\end{figure}

We solve Eq.~\eqref{eq:lindblad} using the open-source code \texttt{QTraj}~\cite{Omar:2021kra} 
choosing for $V_s^{\textrm{pert}}$ and $V_o^{\textrm{pert}}$ the three-loop potentials $V^\text{3L}_s$ and $V^\text{3L}_o$.
For the medium evolution, we include input from 3+1D quasiparticle anisotropic hydrodynamics simulations~\cite{Alqahtani:2020paa}. 
We initiate the coupling to the hydrodynamic evolution at $\tau_\text{med}=\SI{0.6}{\femto\meter}$ and calculate the evolution until a final temperature of $T_f=\SI{250}{\mega\eV}$ is reached, which we take as the lowest temperature for which the strict Brownian limit is valid~\cite{Brambilla:2020qwo}. 
For more details on the numerical setup, we refer the reader to Refs.~\cite{Brambilla:2022ynh,Brambilla:2023hkw}. 
Using \texttt{QTraj}, we sample approximately $\num{140000}$ physical trajectories from the hydrodynamics simulation and $\num{30}$ quantum trajectories per physical trajectory.
At the end of the evolution, we project the wave function of the final state onto the bottomonium eigenstates for $\Upsilon(1S)$, $\Upsilon(2S)$, $\Upsilon(3S)$ and $\chi_b(1P)$, $\chi_b(2P)$, which are relevant through feed-down, obtained from solving the Schr\"odinger
equation for the improved potential $V^\text{3L+np}_s$. 
By taking the ratio of the final state overlap to the initial state overlap, we obtain the survival probability of the corresponding state. 
Including the effect of feed-down as  described in Refs.~\cite{Brambilla:2020qwo,Brambilla:2022ynh}, we finally obtain predictions for the bottomonium nuclear modification factor $R_{AA}$, which is the bottomonium yield in heavy-ion collisions normalized by the bottomonium yield in proton-proton collisions.
The results for $R_{AA}$ in terms of the number of participating nucleons, $N_\text{part}$, 
obtained from using the three-loop potentials $V^\text{3L}_s$ and $V^\text{3L}_o$ in the Lindblad equation, are shown in Figure~\ref{fig:resultsvpert}. 

For the numerical computation, we use the transport coefficient values obtained in section~\ref{sec:transp}, i.e., $\hat\gamma=0$, $\gamma(T=0)=\SI{0.017}{\giga\eV\tothe{3}}$ and $\hat\kappa = \num{1.88\pm 0.16}$. 
The bands indicate the uncertainty in $\hat\kappa$ and the statistical uncertainty from the quantum trajectory algorithm. 
The dash-dotted and dashed lines are the curves for $\hat\kappa=\num{1.72}$ and $\hat\kappa=\num{2.04}$, respectively.
To assess the compatibility with the data, we calculate the weighted relative mean square error (WMSE)
\begin{equation}
    \Delta^2 = \frac{1}{N_\text{data}}\sum_{i}\frac{\left(\frac{R^\text{exp}_{AA,i}-R^\text{QTraj}_{AA,i}}{R^\text{exp}_{AA,i}}\right)^2}{\left(\sigma^\text{exp}_i\big/R^\text{exp}_{AA,i}\right)^2+\left(\sigma^\text{QTraj}_i\big/R^\text{QTraj}_{AA,i}\right)^2}  \, ,
\end{equation}
where the sum runs over the discrete values of $N_{\rm part}$ corresponding to the different measurements by the experiments, 
$R^\text{exp}_{AA,i}$ denotes the $i$th measurement, $R^\text{QTraj}_{AA,i}$ the corresponding prediction from \texttt{QTraj}, and $\sigma^\text{exp}_i$ and $\sigma^\text{QTraj}_i$ the relative uncertainties. 
The data are measurements by the ALICE~\cite{ALICE:2020wwx}, ATLAS~\cite{ATLAS:2022exb}, and CMS~\cite{CMS:2018zza,CMS:2023lfu} collaborations. 
The \texttt{QTraj} predictions and corresponding uncertainties at $N_\text{part}$ given by the experiments are obtained by linear extrapolation between \texttt{QTraj} prediction points. 
The WMSE for the three-loop potentials with the above parameters is $\Delta^2 = \num{2.1e-3}$. 
In contrast, results obtained by solving the Lindblad equation with $V_s^{\textrm{pert}}$ and $V_o^{\textrm{pert}}$ set to the Coulomb potentials $V^c_s$ and $V^c_o$ with $\hat\gamma=0$, $\gamma(T=0)=0$ and $\hat\kappa=\num{0.33}$ 
and projecting onto bottomonium eigenstates obtained by solving the Schrödinger equation for the Coulomb potential $V^c_s$ give $\Delta^2 = \num{0.22}$, 
showing that the three-loop potential is better suited to describe the data than the simple Coulomb potential.
We note that the wave functions calculated with $V_s^c$ and $V^\text{3L+np}_s$, especially the wave functions of the $2S$ and $3S$ states, differ considerably due to the steeper slope of the $V^\text{3L+np}_s$ potential at large distances.

Finally, we remark that the results with the Coulomb potential of Refs.~\cite{Brambilla:2020qwo,Brambilla:2021wkt} were obtained using different parameters for the $b$-mass and $\alpha_s$ than here and, importantly, different values for the transport coefficients.
In this work, however, we constrain the transport coefficients to reproduce the thermal widths of the $S$- and $P$-wave states recently computed in lattice QCD.
Under this constraint, the three-loop potential provides a better description of the bottomonium nuclear modification factor data from the LHC experiments.

\section{Conclusions}
\label{sec:conc}

In the paper, we compute bottomonium suppression in heavy-ion collisions from an evolution equation that encompasses 
the three-loop quarkonium potential and the vacuum part of the dispersive transport coefficient $\gamma(T=0)$. 
There are a number of advantages to this approach.
First, the three-loop potential plus the addition of a small nonperturbative correction derived from the operator product expansion (with the bottom mass fixed on the $1S$ state) provides a realistic description of the spin-averaged bottomonium spectrum; see Table~\ref{tab:spectrum}.
Second, the thermal decay width that follows from the bottomonium wave functions is consistent with that computed in lattice QCD; this allows to fit reasonably well the quarkonium momentum diffusion coefficient on the lattice data;
see Fig.~\ref{fig:width}.
Finally, the obtained bottomonium nuclear modification factor shown in Fig.~\ref{fig:resultsvpert} provides a more accurate description of the LHC data in the strict Brownian limit 
than previous analyses with the Coulomb potential. 
In the future, this study could be extended beyond the Brownian limit by including $E/(\pi T)$ corrections,
as done for the Coulomb potential in Ref.~\cite{Brambilla:2022ynh}.

\section*{Acknowledgments}
N.B., T.M., and A.V. acknowledge support by the DFG cluster of excellence ORIGINS funded by the Deutsche Forschungsgemeinschaft (DFG) under Germany's Excellence Strategy - EXC-2094-390783311. The work of N.B. is supported by the DFG Grant No. BR 4058/5-1 ``Open Quantum Sytems and Effective Field Theories for hard probes of hot and/or dense medium''. N.B., T.M., and A.V. acknowledge support from STRONG-2020, European Union's Horizon 2020 research and innovation program under grant agreement No. 824093.  M.S.\ was supported by the U.S.\ Department of Energy, Office of Science, Office of Nuclear Physics (Nuclear Theory) under contract number DE-SC0013470.
P.V.G. was supported by the U.S. Department of Energy
Award No.~DE-SC0019095.
P.V.G. is grateful for the support and hospitality of the Fermilab theory group.
Fermilab is operated by Fermi Research Alliance, LLC under Contract No.~DE-AC02-07CH11359 with the United States Department of Energy.

\appendix
\section{Potentials}
\label{app:pot}

For convenience, we provide the explicit expressions of the coefficients $V^{(k)}_{s,\text{RS}^\prime}$ and $V^{(k)}_{o,\text{RS}^\prime}$ used in the potentials \eqref{eq:singlet-pert} and \eqref{eq:octet}.
In the singlet case, we have
\begin{equation}
V^{(0)}_{s,\text{RS}^\prime}=-\frac{C_F}{r} \, , 
\end{equation}
\begin{equation}
V^{(1)}_{s,\text{RS}^\prime}=-\frac{C_F}{4\pi r}\left[a_1 + 2\log(r\nu e^{\gamma_E})\beta_0\right] + 2\delta m^{(1)}_{\text{RS}^\prime} \, ,
\end{equation}
\begin{equation}
V^{(2)}_{s,\text{RS}^\prime}=-\frac{C_F}{(4\pi)^2 r}\left[(a_2 + \log(r\nu e^{\gamma_E})(4a_1\beta_0 + 2\beta_1)) +\left(\frac{\pi^2}{3} + 4\log^2(r\nu e^{\gamma_E})\right)\beta^2_0\right] + 2\delta m^{(2)}_{\text{RS}^\prime} \, ,
\end{equation}
\begin{align}
\begin{split}
V^{(3)}_{s,\text{RS}^\prime}=&-\frac{C_F}{(4\pi)^3 r}\bigg[(a_3 + a_1\beta^2_0\pi^2 + \frac{5\pi^2}{6}\beta_0\beta_1+16\zeta^3(3)\beta^3_0 \\ &+ (2\pi^2\beta^3_0+6a_2\beta_0 + 4a_1\beta_1 + 2\beta_2+144\pi^2)\log(r\nu e^{\gamma_E}) \\ &+ (12a_1\beta^2_0 + 10\beta_0\beta_1)\log^2(r\nu e^{\gamma_E}) + 8\beta^3_0\log^3(r\nu e^{\gamma_E})+144\pi^2\log\frac{\nu_{\textrm{us}}}{\nu}\bigg] + 2\delta m^{(3)}_{\text{RS}^\prime} \, ,
\end{split}
\end{align}
where $\beta_i$ are the coefficients of the $\beta$-function. 
We work in the convention where $\beta_0 = 11 C_A/3 -4 T_F n_f/3$, 
with $n_f$ the number of massless flavors, which we set to $n_f=3$ in our bottomonium calculations~\cite{Brambilla:2001qk}. 
The scale $\nu$ is the renormalization scale and $\nu_{\text{us}}$ the factorization scale separating the soft from the ultrasoft region (see Sec.~\ref{sec:pot} for discussion of the scales). The SU($N$) Casimirs are $C_F = T_F (N_c^2-1)/N_c$, $C_A = 2T_F N_c$ with $N_c$ the number of colors and $T_F=1/2$.
The expressions for $a_1, a_2$, and $a_3$ were computed in~Refs.~\cite{Fischler:1977yf,Billoire:1979ih}, Refs.~\cite{Peter:1997me,Schroder:1998vy} and~Refs.~\cite{Anzai:2009tm,Smirnov:2009fh}, respectively. The ultrasoft correction in $V^{(3)}_{s,\text{RS}^\prime}$ was computed in Ref.~\cite{Brambilla:1999qa}. 
The renormalon contributions $\delta m_{\text{RS}^\prime}$ are given by~\cite{Pineda:2001zq}
\begin{equation}
\delta m_{\text{RS}^\prime}=\nu_f\sum^3_{k=0}\delta m^{(k)}_{\text{RS}^\prime}\left(\frac{\nu_f}{\nu}\right)\alpha^{k+1}_s(\nu) \, ,
\end{equation}
with
\begin{align}
& \delta m_{\mathrm{RS}^\prime}^{(0)}=0, \qquad \delta m_{\mathrm{RS}^\prime}^{(1)}\left(\frac{\nu_f}{\nu}\right)=N_m \frac{\beta_0}{2 \pi} S(1, b), \\
& \delta m_{\mathrm{RS}^\prime}^{(2)}\left(\frac{\nu_f}{\nu}\right)=N_m\left(\frac{\beta_0}{2 \pi}\right)\left[S(1, b) \frac{2 d_0\left(\nu, \nu_f\right)}{\pi}+\left(\frac{\beta_0}{2 \pi}\right) S(2, b)\right], \\
\begin{split}
& \delta m_{\mathrm{RS}^\prime}^{(3)}\left(\frac{\nu_f}{\nu}\right)=N_m\left(\frac{\beta_0}{2 \pi}\right)\bigg[S(1, b) \frac{3 d_0^2\left(\nu, \nu_f\right)+2 d_1\left(\nu, \nu_f\right)}{\pi^2}+\left(\frac{\beta_0}{2 \pi}\right) S(2, b) \frac{3 d_0\left(\nu, \nu_f\right)}{\pi}\\&\qquad\qquad\qquad\qquad\qquad\qquad+\left(\frac{\beta_0}{2 \pi}\right)^2 S(3, b)\bigg] , 
\end{split}
\end{align}
where
\begin{equation}
S(n, b)=\sum_{k=0}^2 c_k \frac{\Gamma(n+1+b-k)}{\Gamma(1+b-k)}, \quad d_k\left(\nu, \nu_f\right)=\frac{\beta_k}{2^{1+2 k}} \ln \left(\frac{\nu}{\nu_f}\right),
\end{equation}
and
\begin{equation}
c_0=1, \quad c_1=\frac{\beta_1^2-\beta_0 \beta_2}{4 \beta_0^4 b}, \quad c_2=\frac{\beta_1^4+4 \beta_0^3 \beta_1 \beta_2-2 \beta_0 \beta_1^2 \beta_2+\beta_0^2\left(-2 \beta_1^3+\beta_2^2\right)-2 \beta_0^4 \beta_3}{32 \beta_0^8 b(b-1)} \, .
\end{equation}
The scale $\nu_f$ is the renormalon factorization scale and $b=\beta_1/(2\beta^2_0)$. 
For $N_m$ we use $N_m = 0.563126$ from~Ref.~\cite{Ayala:2014yxa}. 
There are also other determinations, e.g., $N_m = 0.574974$ from Ref.~\cite{Pineda:2001zq}, and $N_m = 0.535$ from Ref.~\cite{Komijani:2017vep}, which only slightly alter the spectrum and the bottom mass.
 
For the octet potential, we use the three-loop one from~Ref.~\cite{Anzai:2013tja}.
The coefficients $V^{(k)}_{o,\text{RS}}$ read
\begin{equation}
V^{(0)}_{o,\text{RS}^\prime}=-\frac{C^{[8]}}{r} \, ,
\end{equation}
\begin{equation}
V^{(1)}_{o,\text{RS}^\prime}=-\frac{C^{[8]}}{4\pi r}\left[a^{[8]}_1 + 2\beta_0 \log(r\nu e^{\gamma_E})\right]-\delta V^{(1)}_{o,\text{RS}^\prime} \, ,
\end{equation}
\begin{equation}
V^{(2)}_{o,\text{RS}^\prime}=-\frac{C^{[8]}}{16\pi^2 r}\left[a^{[8]}_2 + \beta^2_0\left(\frac{\pi^2}{3} + 4\log^2(r\nu e^{\gamma_E})\right) + \log(r\nu e^{\gamma_E})(4a^{[8]}_1\beta_0+2\beta_1)\right]-\delta V^{(2)}_{o,\text{RS}^\prime} \, ,
\end{equation}
\begin{align}
 \begin{split}
    V^{(3)}_{o,\text{RS}^\prime}=-&\frac{C^{[8]}}{64\pi^2 r}\bigg[a^{[8]}_3 + 8\beta^3_0\big(\log^3(r\nu e^{\gamma_E})+\frac{\pi^2}{4}\log(r\nu e^{\gamma_E}) +2\zeta(3)\big)\\&+(10\beta_0\beta_1 + 12\beta^2_0a^{[8]}_1)\bigg(\log^2(r\nu e^{\gamma_E}) + \frac{\pi^2}{12}\bigg) \\&+ (2\beta_2+4\beta_1a^{[8]}_1+6\beta_0a^{[8]}_2)\log(r\nu e^{\gamma_E}) + \frac{64\pi^2}{3}N^3_c\log(r\nu)\bigg]-\delta V^{(3)}_{o,\text{RS}^\prime} \, ,
 \end{split}
\end{align}
with $C^{[8]}=C_F-C_A/2$. 
The renormalon contributions $\delta V^{(i)}_{o,\text{RS}^\prime}$ have the same form as $\delta m^{(i)}_{\text{RS}^\prime}$ but with normalization $N_{V_o}$ instead of $N_{m}$, 
which is set to $N_{V_o}=\num {0.114}$~\cite{Bali:2003jq}.
The coefficients $a^{[8]}_1$ and $a^{[8]}_2$ are defined as 
\begin{equation}
a^{[8]}_1 = a_1 \, ,
\end{equation}
\begin{equation}
a^{[8]}_2 = a_2+N^2_c\pi^2(\pi^2-12)  \, ,
\end{equation}
where $a_1$ and $a_2$ are from the singlet potential. The coefficient $a^{[8]}_3$ has the form
\begin{equation}
a^{[8]}_3 = a^{[8],(3)}_3n^3_f+a^{[8],(2)}_3n^2_f+a^{[8],(1)}_3n_f+a^{[8],(0)}_3  \, ,
\end{equation}
with
\begin{equation}
a^{[8],(3)}_3=-\left(\frac{20}{9}\right)^3T^3_F  \, ,
\end{equation}
\begin{equation}
a^{[8],(2)}_3=\bigg(\frac{12541}{243}+\frac{368\zeta(3)}{3}+\frac{64\pi^4}{135}\bigg)C_AT^2_F + \bigg(\frac{14002}{81}-\frac{416\zeta(3)}{3}\bigg)C_FT^2_F \, ,
\end{equation}
\begin{equation}
a^{[8],(1)}_3 = -327.2N^2_c + \frac{66133}{648}-\frac{112\pi^2}{9}-\frac{272\zeta(3)}{3}+\frac{8\pi^4}{3}-\frac{32\pi^2\zeta(3)}{3}+20\zeta(5)-12.6\frac{1}{N^2_c} \, ,
\end{equation}
\begin{equation}
a^{[8],(0)}_3=-17.0N_c + 499.4N^3_c-97.6N^3_c \, .
\end{equation}

\bibliographystyle{apsrev4-1}
\renewcommand*{\bibfont}{\footnotesize}
\bibliography{lit.bib}

\begin{thebibliography}{58}%
\makeatletter
\providecommand \@ifxundefined [1]{%
 \@ifx{#1\undefined}
}%
\providecommand \@ifnum [1]{%
 \ifnum #1\expandafter \@firstoftwo
 \else \expandafter \@secondoftwo
 \fi
}%
\providecommand \@ifx [1]{%
 \ifx #1\expandafter \@firstoftwo
 \else \expandafter \@secondoftwo
 \fi
}%
\providecommand \natexlab [1]{#1}%
\providecommand \enquote  [1]{``#1''}%
\providecommand \bibnamefont  [1]{#1}%
\providecommand \bibfnamefont [1]{#1}%
\providecommand \citenamefont [1]{#1}%
\providecommand \href@noop [0]{\@secondoftwo}%
\providecommand \href [0]{\begingroup \@sanitize@url \@href}%
\providecommand \@href[1]{\@@startlink{#1}\@@href}%
\providecommand \@@href[1]{\endgroup#1\@@endlink}%
\providecommand \@sanitize@url [0]{\catcode `\\12\catcode `\$12\catcode `\&12\catcode `\#12\catcode `\^12\catcode `\_12\catcode `\%12\relax}%
\providecommand \@@startlink[1]{}%
\providecommand \@@endlink[0]{}%
\providecommand \url  [0]{\begingroup\@sanitize@url \@url }%
\providecommand \@url [1]{\endgroup\@href {#1}{\urlprefix }}%
\providecommand \urlprefix  [0]{URL }%
\providecommand \Eprint [0]{\href }%
\providecommand \doibase [0]{http://dx.doi.org/}%
\providecommand \selectlanguage [0]{\@gobble}%
\providecommand \bibinfo  [0]{\@secondoftwo}%
\providecommand \bibfield  [0]{\@secondoftwo}%
\providecommand \translation [1]{[#1]}%
\providecommand \BibitemOpen [0]{}%
\providecommand \bibitemStop [0]{}%
\providecommand \bibitemNoStop [0]{.\EOS\space}%
\providecommand \EOS [0]{\spacefactor3000\relax}%
\providecommand \BibitemShut  [1]{\csname bibitem#1\endcsname}%
\let\auto@bib@innerbib\@empty
\bibitem [{\citenamefont {Matsui}\ and\ \citenamefont {Satz}(1986)}]{Matsui:1986dk}%
  \BibitemOpen
  \bibfield  {author} {\bibinfo {author} {\bibfnamefont {T.}~\bibnamefont {Matsui}}\ and\ \bibinfo {author} {\bibfnamefont {H.}~\bibnamefont {Satz}},\ }\href {\doibase 10.1016/0370-2693(86)91404-8} {\bibfield  {journal} {\bibinfo  {journal} {Phys. Lett. B}\ }\textbf {\bibinfo {volume} {178}},\ \bibinfo {pages} {416} (\bibinfo {year} {1986})}\BibitemShut {NoStop}%
\bibitem [{\citenamefont {Andronic}\ \emph {et~al.}(2024)\citenamefont {Andronic} \emph {et~al.}}]{Andronic:2024oxz}%
  \BibitemOpen
  \bibfield  {author} {\bibinfo {author} {\bibfnamefont {A.}~\bibnamefont {Andronic}} \emph {et~al.},\ }\href@noop {} {\  (\bibinfo {year} {2024})},\ \Eprint {http://arxiv.org/abs/2402.04366} {arXiv:2402.04366 [nucl-th]} \BibitemShut {NoStop}%
\bibitem [{\citenamefont {Brambilla}\ \emph {et~al.}(2005)\citenamefont {Brambilla}, \citenamefont {Pineda}, \citenamefont {Soto},\ and\ \citenamefont {Vairo}}]{Brambilla:2004jw}%
  \BibitemOpen
  \bibfield  {author} {\bibinfo {author} {\bibfnamefont {N.}~\bibnamefont {Brambilla}}, \bibinfo {author} {\bibfnamefont {A.}~\bibnamefont {Pineda}}, \bibinfo {author} {\bibfnamefont {J.}~\bibnamefont {Soto}}, \ and\ \bibinfo {author} {\bibfnamefont {A.}~\bibnamefont {Vairo}},\ }\href {\doibase 10.1103/RevModPhys.77.1423} {\bibfield  {journal} {\bibinfo  {journal} {Rev. Mod. Phys.}\ }\textbf {\bibinfo {volume} {77}},\ \bibinfo {pages} {1423} (\bibinfo {year} {2005})},\ \Eprint {http://arxiv.org/abs/hep-ph/0410047} {arXiv:hep-ph/0410047} \BibitemShut {NoStop}%
\bibitem [{\citenamefont {Brambilla}\ \emph {et~al.}(2017)\citenamefont {Brambilla}, \citenamefont {Escobedo}, \citenamefont {Soto},\ and\ \citenamefont {Vairo}}]{Brambilla:2016wgg}%
  \BibitemOpen
  \bibfield  {author} {\bibinfo {author} {\bibfnamefont {N.}~\bibnamefont {Brambilla}}, \bibinfo {author} {\bibfnamefont {M.~A.}\ \bibnamefont {Escobedo}}, \bibinfo {author} {\bibfnamefont {J.}~\bibnamefont {Soto}}, \ and\ \bibinfo {author} {\bibfnamefont {A.}~\bibnamefont {Vairo}},\ }\href {\doibase 10.1103/PhysRevD.96.034021} {\bibfield  {journal} {\bibinfo  {journal} {Phys. Rev. D}\ }\textbf {\bibinfo {volume} {96}},\ \bibinfo {pages} {034021} (\bibinfo {year} {2017})},\ \Eprint {http://arxiv.org/abs/1612.07248} {arXiv:1612.07248} \BibitemShut {NoStop}%
\bibitem [{\citenamefont {Brambilla}\ \emph {et~al.}(2018)\citenamefont {Brambilla}, \citenamefont {Escobedo}, \citenamefont {Soto},\ and\ \citenamefont {Vairo}}]{Brambilla:2017zei}%
  \BibitemOpen
  \bibfield  {author} {\bibinfo {author} {\bibfnamefont {N.}~\bibnamefont {Brambilla}}, \bibinfo {author} {\bibfnamefont {M.~A.}\ \bibnamefont {Escobedo}}, \bibinfo {author} {\bibfnamefont {J.}~\bibnamefont {Soto}}, \ and\ \bibinfo {author} {\bibfnamefont {A.}~\bibnamefont {Vairo}},\ }\href {\doibase 10.1103/PhysRevD.97.074009} {\bibfield  {journal} {\bibinfo  {journal} {Phys. Rev. D}\ }\textbf {\bibinfo {volume} {97}},\ \bibinfo {pages} {074009} (\bibinfo {year} {2018})},\ \Eprint {http://arxiv.org/abs/1711.04515} {arXiv:1711.04515} \BibitemShut {NoStop}%
\bibitem [{\citenamefont {Brambilla}\ \emph {et~al.}(2021{\natexlab{a}})\citenamefont {Brambilla}, \citenamefont {Escobedo}, \citenamefont {Strickland}, \citenamefont {Vairo}, \citenamefont {Vander~Griend},\ and\ \citenamefont {Weber}}]{Brambilla:2020qwo}%
  \BibitemOpen
  \bibfield  {author} {\bibinfo {author} {\bibfnamefont {N.}~\bibnamefont {Brambilla}}, \bibinfo {author} {\bibfnamefont {M.~A.}\ \bibnamefont {Escobedo}}, \bibinfo {author} {\bibfnamefont {M.}~\bibnamefont {Strickland}}, \bibinfo {author} {\bibfnamefont {A.}~\bibnamefont {Vairo}}, \bibinfo {author} {\bibfnamefont {P.}~\bibnamefont {Vander~Griend}}, \ and\ \bibinfo {author} {\bibfnamefont {J.~H.}\ \bibnamefont {Weber}},\ }\href {\doibase 10.1007/JHEP05(2021)136} {\bibfield  {journal} {\bibinfo  {journal} {JHEP}\ }\textbf {\bibinfo {volume} {05}},\ \bibinfo {pages} {136} (\bibinfo {year} {2021}{\natexlab{a}})},\ \Eprint {http://arxiv.org/abs/2012.01240} {arXiv:2012.01240} \BibitemShut {NoStop}%
\bibitem [{\citenamefont {Akamatsu}(2022)}]{Akamatsu:2020ypb}%
  \BibitemOpen
  \bibfield  {author} {\bibinfo {author} {\bibfnamefont {Y.}~\bibnamefont {Akamatsu}},\ }\href {\doibase 10.1016/j.ppnp.2021.103932} {\bibfield  {journal} {\bibinfo  {journal} {Prog. Part. Nucl. Phys.}\ }\textbf {\bibinfo {volume} {123}},\ \bibinfo {pages} {103932} (\bibinfo {year} {2022})},\ \Eprint {http://arxiv.org/abs/2009.10559} {arXiv:2009.10559 [nucl-th]} \BibitemShut {NoStop}%
\bibitem [{\citenamefont {Yao}(2021)}]{Yao:2021lus}%
  \BibitemOpen
  \bibfield  {author} {\bibinfo {author} {\bibfnamefont {X.}~\bibnamefont {Yao}},\ }\href {\doibase 10.1142/S0217751X21300106} {\bibfield  {journal} {\bibinfo  {journal} {Int. J. Mod. Phys. A}\ }\textbf {\bibinfo {volume} {36}},\ \bibinfo {pages} {2130010} (\bibinfo {year} {2021})},\ \Eprint {http://arxiv.org/abs/2102.01736} {arXiv:2102.01736 [hep-ph]} \BibitemShut {NoStop}%
\bibitem [{\citenamefont {Miura}\ \emph {et~al.}(2022)\citenamefont {Miura}, \citenamefont {Akamatsu}, \citenamefont {Asakawa},\ and\ \citenamefont {Kaida}}]{Miura:2022arv}%
  \BibitemOpen
  \bibfield  {author} {\bibinfo {author} {\bibfnamefont {T.}~\bibnamefont {Miura}}, \bibinfo {author} {\bibfnamefont {Y.}~\bibnamefont {Akamatsu}}, \bibinfo {author} {\bibfnamefont {M.}~\bibnamefont {Asakawa}}, \ and\ \bibinfo {author} {\bibfnamefont {Y.}~\bibnamefont {Kaida}},\ }\href {\doibase 10.1103/PhysRevD.106.074001} {\bibfield  {journal} {\bibinfo  {journal} {Phys. Rev. D}\ }\textbf {\bibinfo {volume} {106}},\ \bibinfo {pages} {074001} (\bibinfo {year} {2022})},\ \Eprint {http://arxiv.org/abs/2205.15551} {arXiv:2205.15551 [nucl-th]} \BibitemShut {NoStop}%
\bibitem [{\citenamefont {Brambilla}\ \emph {et~al.}(2022)\citenamefont {Brambilla}, \citenamefont {Escobedo}, \citenamefont {Islam}, \citenamefont {Strickland}, \citenamefont {Tiwari}, \citenamefont {Vairo},\ and\ \citenamefont {Vander~Griend}}]{Brambilla:2022ynh}%
  \BibitemOpen
  \bibfield  {author} {\bibinfo {author} {\bibfnamefont {N.}~\bibnamefont {Brambilla}}, \bibinfo {author} {\bibfnamefont {M.~A.}\ \bibnamefont {Escobedo}}, \bibinfo {author} {\bibfnamefont {A.}~\bibnamefont {Islam}}, \bibinfo {author} {\bibfnamefont {M.}~\bibnamefont {Strickland}}, \bibinfo {author} {\bibfnamefont {A.}~\bibnamefont {Tiwari}}, \bibinfo {author} {\bibfnamefont {A.}~\bibnamefont {Vairo}}, \ and\ \bibinfo {author} {\bibfnamefont {P.}~\bibnamefont {Vander~Griend}},\ }\href {\doibase 10.1007/JHEP08(2022)303} {\bibfield  {journal} {\bibinfo  {journal} {JHEP}\ }\textbf {\bibinfo {volume} {08}},\ \bibinfo {pages} {303} (\bibinfo {year} {2022})},\ \Eprint {http://arxiv.org/abs/2205.10289} {arXiv:2205.10289} \BibitemShut {NoStop}%
\bibitem [{\citenamefont {Brambilla}\ \emph {et~al.}(2023)\citenamefont {Brambilla}, \citenamefont {Escobedo}, \citenamefont {Islam}, \citenamefont {Strickland}, \citenamefont {Tiwari}, \citenamefont {Vairo},\ and\ \citenamefont {Vander~Griend}}]{Brambilla:2023hkw}%
  \BibitemOpen
  \bibfield  {author} {\bibinfo {author} {\bibfnamefont {N.}~\bibnamefont {Brambilla}}, \bibinfo {author} {\bibfnamefont {M.~A.}\ \bibnamefont {Escobedo}}, \bibinfo {author} {\bibfnamefont {A.}~\bibnamefont {Islam}}, \bibinfo {author} {\bibfnamefont {M.}~\bibnamefont {Strickland}}, \bibinfo {author} {\bibfnamefont {A.}~\bibnamefont {Tiwari}}, \bibinfo {author} {\bibfnamefont {A.}~\bibnamefont {Vairo}}, \ and\ \bibinfo {author} {\bibfnamefont {P.}~\bibnamefont {Vander~Griend}},\ }\href {\doibase 10.1103/PhysRevD.108.L011502} {\bibfield  {journal} {\bibinfo  {journal} {Phys. Rev. D}\ }\textbf {\bibinfo {volume} {108}},\ \bibinfo {pages} {L011502} (\bibinfo {year} {2023})},\ \Eprint {http://arxiv.org/abs/2302.11826} {arXiv:2302.11826} \BibitemShut {NoStop}%
\bibitem [{\citenamefont {Acharya}\ \emph {et~al.}(2021)\citenamefont {Acharya} \emph {et~al.}}]{ALICE:2020wwx}%
  \BibitemOpen
  \bibfield  {author} {\bibinfo {author} {\bibfnamefont {S.}~\bibnamefont {Acharya}} \emph {et~al.} (\bibinfo {collaboration} {ALICE}),\ }\href {\doibase 10.1016/j.physletb.2021.136579} {\bibfield  {journal} {\bibinfo  {journal} {Phys. Lett. B}\ }\textbf {\bibinfo {volume} {822}},\ \bibinfo {pages} {136579} (\bibinfo {year} {2021})},\ \Eprint {http://arxiv.org/abs/2011.05758} {arXiv:2011.05758} \BibitemShut {NoStop}%
\bibitem [{\citenamefont {Aad}\ \emph {et~al.}(2023)\citenamefont {Aad} \emph {et~al.}}]{ATLAS:2022exb}%
  \BibitemOpen
  \bibfield  {author} {\bibinfo {author} {\bibfnamefont {G.}~\bibnamefont {Aad}} \emph {et~al.} (\bibinfo {collaboration} {ATLAS}),\ }\href {\doibase 10.1103/PhysRevC.107.054912} {\bibfield  {journal} {\bibinfo  {journal} {Phys. Rev. C}\ }\textbf {\bibinfo {volume} {107}},\ \bibinfo {pages} {054912} (\bibinfo {year} {2023})},\ \Eprint {http://arxiv.org/abs/2205.03042} {arXiv:2205.03042} \BibitemShut {NoStop}%
\bibitem [{\citenamefont {Sirunyan}\ \emph {et~al.}(2019)\citenamefont {Sirunyan} \emph {et~al.}}]{CMS:2018zza}%
  \BibitemOpen
  \bibfield  {author} {\bibinfo {author} {\bibfnamefont {A.~M.}\ \bibnamefont {Sirunyan}} \emph {et~al.} (\bibinfo {collaboration} {CMS}),\ }\href {\doibase 10.1016/j.physletb.2019.01.006} {\bibfield  {journal} {\bibinfo  {journal} {Phys. Lett. B}\ }\textbf {\bibinfo {volume} {790}},\ \bibinfo {pages} {270} (\bibinfo {year} {2019})},\ \Eprint {http://arxiv.org/abs/1805.09215} {arXiv:1805.09215} \BibitemShut {NoStop}%
\bibitem [{\citenamefont {Tumasyan}\ \emph {et~al.}(2023)\citenamefont {Tumasyan} \emph {et~al.}}]{CMS:2023lfu}%
  \BibitemOpen
  \bibfield  {author} {\bibinfo {author} {\bibfnamefont {A.}~\bibnamefont {Tumasyan}} \emph {et~al.} (\bibinfo {collaboration} {CMS}),\ }\href@noop {} {\  (\bibinfo {year} {2023})},\ \Eprint {http://arxiv.org/abs/2303.17026} {arXiv:2303.17026} \BibitemShut {NoStop}%
\bibitem [{\citenamefont {Caswell}\ and\ \citenamefont {Lepage}(1986)}]{Caswell:1985ui}%
  \BibitemOpen
  \bibfield  {author} {\bibinfo {author} {\bibfnamefont {W.~E.}\ \bibnamefont {Caswell}}\ and\ \bibinfo {author} {\bibfnamefont {G.~P.}\ \bibnamefont {Lepage}},\ }\href {\doibase 10.1016/0370-2693(86)91297-9} {\bibfield  {journal} {\bibinfo  {journal} {Phys. Lett. B}\ }\textbf {\bibinfo {volume} {167}},\ \bibinfo {pages} {437} (\bibinfo {year} {1986})}\BibitemShut {NoStop}%
\bibitem [{\citenamefont {Bodwin}\ \emph {et~al.}(1995)\citenamefont {Bodwin}, \citenamefont {Braaten},\ and\ \citenamefont {Lepage}}]{Bodwin:1994jh}%
  \BibitemOpen
  \bibfield  {author} {\bibinfo {author} {\bibfnamefont {G.~T.}\ \bibnamefont {Bodwin}}, \bibinfo {author} {\bibfnamefont {E.}~\bibnamefont {Braaten}}, \ and\ \bibinfo {author} {\bibfnamefont {G.~P.}\ \bibnamefont {Lepage}},\ }\href {\doibase 10.1103/PhysRevD.55.5853} {\bibfield  {journal} {\bibinfo  {journal} {Phys. Rev. D}\ }\textbf {\bibinfo {volume} {51}},\ \bibinfo {pages} {1125} (\bibinfo {year} {1995})},\ \bibinfo {note} {[Erratum: Phys.Rev.D 55, 5853 (1997)]},\ \Eprint {http://arxiv.org/abs/hep-ph/9407339} {arXiv:hep-ph/9407339} \BibitemShut {NoStop}%
\bibitem [{\citenamefont {Pineda}\ and\ \citenamefont {Soto}(1998)}]{Pineda:1997bj}%
  \BibitemOpen
  \bibfield  {author} {\bibinfo {author} {\bibfnamefont {A.}~\bibnamefont {Pineda}}\ and\ \bibinfo {author} {\bibfnamefont {J.}~\bibnamefont {Soto}},\ }\href {\doibase 10.1016/S0920-5632(97)01102-X} {\bibfield  {journal} {\bibinfo  {journal} {Nucl. Phys. B Proc. Suppl.}\ }\textbf {\bibinfo {volume} {64}},\ \bibinfo {pages} {428} (\bibinfo {year} {1998})},\ \Eprint {http://arxiv.org/abs/hep-ph/9707481} {arXiv:hep-ph/9707481} \BibitemShut {NoStop}%
\bibitem [{\citenamefont {Brambilla}\ \emph {et~al.}(2000)\citenamefont {Brambilla}, \citenamefont {Pineda}, \citenamefont {Soto},\ and\ \citenamefont {Vairo}}]{Brambilla:1999xf}%
  \BibitemOpen
  \bibfield  {author} {\bibinfo {author} {\bibfnamefont {N.}~\bibnamefont {Brambilla}}, \bibinfo {author} {\bibfnamefont {A.}~\bibnamefont {Pineda}}, \bibinfo {author} {\bibfnamefont {J.}~\bibnamefont {Soto}}, \ and\ \bibinfo {author} {\bibfnamefont {A.}~\bibnamefont {Vairo}},\ }\href {\doibase 10.1016/S0550-3213(99)00693-8} {\bibfield  {journal} {\bibinfo  {journal} {Nucl. Phys. B}\ }\textbf {\bibinfo {volume} {566}},\ \bibinfo {pages} {275} (\bibinfo {year} {2000})},\ \Eprint {http://arxiv.org/abs/hep-ph/9907240} {arXiv:hep-ph/9907240} \BibitemShut {NoStop}%
\bibitem [{\citenamefont {Brambilla}\ and\ \citenamefont {Vairo}(1999)}]{Brambilla:1999ja}%
  \BibitemOpen
  \bibfield  {author} {\bibinfo {author} {\bibfnamefont {N.}~\bibnamefont {Brambilla}}\ and\ \bibinfo {author} {\bibfnamefont {A.}~\bibnamefont {Vairo}},\ }in\ \href@noop {} {\emph {\bibinfo {booktitle} {{13th Annual HUGS AT CEBAF}}}}\ (\bibinfo {year} {1999})\ pp.\ \bibinfo {pages} {151--220},\ \Eprint {http://arxiv.org/abs/hep-ph/9904330} {arXiv:hep-ph/9904330} \BibitemShut {NoStop}%
\bibitem [{\citenamefont {Brambilla}\ \emph {et~al.}(1999)\citenamefont {Brambilla}, \citenamefont {Pineda}, \citenamefont {Soto},\ and\ \citenamefont {Vairo}}]{Brambilla:1999qa}%
  \BibitemOpen
  \bibfield  {author} {\bibinfo {author} {\bibfnamefont {N.}~\bibnamefont {Brambilla}}, \bibinfo {author} {\bibfnamefont {A.}~\bibnamefont {Pineda}}, \bibinfo {author} {\bibfnamefont {J.}~\bibnamefont {Soto}}, \ and\ \bibinfo {author} {\bibfnamefont {A.}~\bibnamefont {Vairo}},\ }\href {\doibase 10.1103/PhysRevD.60.091502} {\bibfield  {journal} {\bibinfo  {journal} {Phys. Rev. D}\ }\textbf {\bibinfo {volume} {60}},\ \bibinfo {pages} {091502} (\bibinfo {year} {1999})},\ \Eprint {http://arxiv.org/abs/hep-ph/9903355} {arXiv:hep-ph/9903355} \BibitemShut {NoStop}%
\bibitem [{\citenamefont {Anzai}\ \emph {et~al.}(2010)\citenamefont {Anzai}, \citenamefont {Kiyo},\ and\ \citenamefont {Sumino}}]{Anzai:2009tm}%
  \BibitemOpen
  \bibfield  {author} {\bibinfo {author} {\bibfnamefont {C.}~\bibnamefont {Anzai}}, \bibinfo {author} {\bibfnamefont {Y.}~\bibnamefont {Kiyo}}, \ and\ \bibinfo {author} {\bibfnamefont {Y.}~\bibnamefont {Sumino}},\ }\href {\doibase 10.1103/PhysRevLett.104.112003} {\bibfield  {journal} {\bibinfo  {journal} {Phys. Rev. Lett.}\ }\textbf {\bibinfo {volume} {104}},\ \bibinfo {pages} {112003} (\bibinfo {year} {2010})},\ \Eprint {http://arxiv.org/abs/0911.4335} {arXiv:0911.4335 [hep-ph]} \BibitemShut {NoStop}%
\bibitem [{\citenamefont {Smirnov}\ \emph {et~al.}(2010)\citenamefont {Smirnov}, \citenamefont {Smirnov},\ and\ \citenamefont {Steinhauser}}]{Smirnov:2009fh}%
  \BibitemOpen
  \bibfield  {author} {\bibinfo {author} {\bibfnamefont {A.~V.}\ \bibnamefont {Smirnov}}, \bibinfo {author} {\bibfnamefont {V.~A.}\ \bibnamefont {Smirnov}}, \ and\ \bibinfo {author} {\bibfnamefont {M.}~\bibnamefont {Steinhauser}},\ }\href {\doibase 10.1103/PhysRevLett.104.112002} {\bibfield  {journal} {\bibinfo  {journal} {Phys. Rev. Lett.}\ }\textbf {\bibinfo {volume} {104}},\ \bibinfo {pages} {112002} (\bibinfo {year} {2010})},\ \Eprint {http://arxiv.org/abs/0911.4742} {arXiv:0911.4742 [hep-ph]} \BibitemShut {NoStop}%
\bibitem [{\citenamefont {Pineda}(2001)}]{Pineda:2001zq}%
  \BibitemOpen
  \bibfield  {author} {\bibinfo {author} {\bibfnamefont {A.}~\bibnamefont {Pineda}},\ }\href {\doibase 10.1088/1126-6708/2001/06/022} {\bibfield  {journal} {\bibinfo  {journal} {JHEP}\ }\textbf {\bibinfo {volume} {06}},\ \bibinfo {pages} {022} (\bibinfo {year} {2001})},\ \Eprint {http://arxiv.org/abs/hep-ph/0105008} {arXiv:hep-ph/0105008} \BibitemShut {NoStop}%
\bibitem [{\citenamefont {Brambilla}\ \emph {et~al.}(2001)\citenamefont {Brambilla}, \citenamefont {Sumino},\ and\ \citenamefont {Vairo}}]{Brambilla:2001fw}%
  \BibitemOpen
  \bibfield  {author} {\bibinfo {author} {\bibfnamefont {N.}~\bibnamefont {Brambilla}}, \bibinfo {author} {\bibfnamefont {Y.}~\bibnamefont {Sumino}}, \ and\ \bibinfo {author} {\bibfnamefont {A.}~\bibnamefont {Vairo}},\ }\href {\doibase 10.1016/S0370-2693(01)00611-6} {\bibfield  {journal} {\bibinfo  {journal} {Phys. Lett. B}\ }\textbf {\bibinfo {volume} {513}},\ \bibinfo {pages} {381} (\bibinfo {year} {2001})},\ \Eprint {http://arxiv.org/abs/hep-ph/0101305} {arXiv:hep-ph/0101305} \BibitemShut {NoStop}%
\bibitem [{\citenamefont {Sumino}(2016)}]{Sumino:2016sxe}%
  \BibitemOpen
  \bibfield  {author} {\bibinfo {author} {\bibfnamefont {Y.}~\bibnamefont {Sumino}},\ }\href {\doibase 10.22323/1.260.0011} {\bibfield  {journal} {\bibinfo  {journal} {PoS}\ }\textbf {\bibinfo {volume} {LL2016}},\ \bibinfo {pages} {011} (\bibinfo {year} {2016})},\ \Eprint {http://arxiv.org/abs/1607.03469} {arXiv:1607.03469 [hep-ph]} \BibitemShut {NoStop}%
\bibitem [{\citenamefont {Peset}\ \emph {et~al.}(2018)\citenamefont {Peset}, \citenamefont {Pineda},\ and\ \citenamefont {Segovia}}]{Peset:2018jkf}%
  \BibitemOpen
  \bibfield  {author} {\bibinfo {author} {\bibfnamefont {C.}~\bibnamefont {Peset}}, \bibinfo {author} {\bibfnamefont {A.}~\bibnamefont {Pineda}}, \ and\ \bibinfo {author} {\bibfnamefont {J.}~\bibnamefont {Segovia}},\ }\href {\doibase 10.1103/PhysRevD.98.094003} {\bibfield  {journal} {\bibinfo  {journal} {Phys. Rev. D}\ }\textbf {\bibinfo {volume} {98}},\ \bibinfo {pages} {094003} (\bibinfo {year} {2018})},\ \Eprint {http://arxiv.org/abs/1809.09124} {arXiv:1809.09124 [hep-ph]} \BibitemShut {NoStop}%
\bibitem [{\citenamefont {Kiyo}\ \emph {et~al.}(2010)\citenamefont {Kiyo}, \citenamefont {Pineda},\ and\ \citenamefont {Signer}}]{Kiyo:2010jm}%
  \BibitemOpen
  \bibfield  {author} {\bibinfo {author} {\bibfnamefont {Y.}~\bibnamefont {Kiyo}}, \bibinfo {author} {\bibfnamefont {A.}~\bibnamefont {Pineda}}, \ and\ \bibinfo {author} {\bibfnamefont {A.}~\bibnamefont {Signer}},\ }\href {\doibase 10.1016/j.nuclphysb.2010.08.007} {\bibfield  {journal} {\bibinfo  {journal} {Nucl. Phys. B}\ }\textbf {\bibinfo {volume} {841}},\ \bibinfo {pages} {231} (\bibinfo {year} {2010})},\ \Eprint {http://arxiv.org/abs/1006.2685} {arXiv:1006.2685 [hep-ph]} \BibitemShut {NoStop}%
\bibitem [{\citenamefont {Pineda}\ and\ \citenamefont {Segovia}(2013)}]{Pineda:2013lta}%
  \BibitemOpen
  \bibfield  {author} {\bibinfo {author} {\bibfnamefont {A.}~\bibnamefont {Pineda}}\ and\ \bibinfo {author} {\bibfnamefont {J.}~\bibnamefont {Segovia}},\ }\href {\doibase 10.1103/PhysRevD.87.074024} {\bibfield  {journal} {\bibinfo  {journal} {Phys. Rev. D}\ }\textbf {\bibinfo {volume} {87}},\ \bibinfo {pages} {074024} (\bibinfo {year} {2013})},\ \Eprint {http://arxiv.org/abs/1302.3528} {arXiv:1302.3528 [hep-ph]} \BibitemShut {NoStop}%
\bibitem [{\citenamefont {Segovia}\ \emph {et~al.}(2019)\citenamefont {Segovia}, \citenamefont {Steinbei\ss{}er},\ and\ \citenamefont {Vairo}}]{Segovia:2018qzb}%
  \BibitemOpen
  \bibfield  {author} {\bibinfo {author} {\bibfnamefont {J.}~\bibnamefont {Segovia}}, \bibinfo {author} {\bibfnamefont {S.}~\bibnamefont {Steinbei\ss{}er}}, \ and\ \bibinfo {author} {\bibfnamefont {A.}~\bibnamefont {Vairo}},\ }\href {\doibase 10.1103/PhysRevD.99.074011} {\bibfield  {journal} {\bibinfo  {journal} {Phys. Rev. D}\ }\textbf {\bibinfo {volume} {99}},\ \bibinfo {pages} {074011} (\bibinfo {year} {2019})},\ \Eprint {http://arxiv.org/abs/1811.07590} {arXiv:1811.07590 [hep-ph]} \BibitemShut {NoStop}%
\bibitem [{\citenamefont {Titard}\ and\ \citenamefont {Yndurain}(1994)}]{Titard:1993nn}%
  \BibitemOpen
  \bibfield  {author} {\bibinfo {author} {\bibfnamefont {S.}~\bibnamefont {Titard}}\ and\ \bibinfo {author} {\bibfnamefont {F.~J.}\ \bibnamefont {Yndurain}},\ }\href {\doibase 10.1103/PhysRevD.49.6007} {\bibfield  {journal} {\bibinfo  {journal} {Phys. Rev. D}\ }\textbf {\bibinfo {volume} {49}},\ \bibinfo {pages} {6007} (\bibinfo {year} {1994})},\ \Eprint {http://arxiv.org/abs/hep-ph/9310236} {arXiv:hep-ph/9310236} \BibitemShut {NoStop}%
\bibitem [{\citenamefont {Brambilla}\ and\ \citenamefont {Vairo}(2000)}]{Brambilla:2000am}%
  \BibitemOpen
  \bibfield  {author} {\bibinfo {author} {\bibfnamefont {N.}~\bibnamefont {Brambilla}}\ and\ \bibinfo {author} {\bibfnamefont {A.}~\bibnamefont {Vairo}},\ }in\ \href@noop {} {\emph {\bibinfo {booktitle} {{5th Workshop on QCD (QCD 2000)}}}}\ (\bibinfo {year} {2000})\ pp.\ \bibinfo {pages} {8--16},\ \Eprint {http://arxiv.org/abs/hep-ph/0004192} {arXiv:hep-ph/0004192} \BibitemShut {NoStop}%
\bibitem [{\citenamefont {Larsen}\ \emph {et~al.}(2019)\citenamefont {Larsen}, \citenamefont {Meinel}, \citenamefont {Mukherjee},\ and\ \citenamefont {Petreczky}}]{Larsen:2019bwy}%
  \BibitemOpen
  \bibfield  {author} {\bibinfo {author} {\bibfnamefont {R.}~\bibnamefont {Larsen}}, \bibinfo {author} {\bibfnamefont {S.}~\bibnamefont {Meinel}}, \bibinfo {author} {\bibfnamefont {S.}~\bibnamefont {Mukherjee}}, \ and\ \bibinfo {author} {\bibfnamefont {P.}~\bibnamefont {Petreczky}},\ }\href {\doibase 10.1103/PhysRevD.100.074506} {\bibfield  {journal} {\bibinfo  {journal} {Phys. Rev. D}\ }\textbf {\bibinfo {volume} {100}},\ \bibinfo {pages} {074506} (\bibinfo {year} {2019})},\ \Eprint {http://arxiv.org/abs/1908.08437} {arXiv:1908.08437 [hep-lat]} \BibitemShut {NoStop}%
\bibitem [{\citenamefont {Omar}\ \emph {et~al.}(2022)\citenamefont {Omar}, \citenamefont {Escobedo}, \citenamefont {Islam}, \citenamefont {Strickland}, \citenamefont {Thapa}, \citenamefont {Vander~Griend},\ and\ \citenamefont {Weber}}]{Omar:2021kra}%
  \BibitemOpen
  \bibfield  {author} {\bibinfo {author} {\bibfnamefont {H.~B.}\ \bibnamefont {Omar}}, \bibinfo {author} {\bibfnamefont {M.~A.}\ \bibnamefont {Escobedo}}, \bibinfo {author} {\bibfnamefont {A.}~\bibnamefont {Islam}}, \bibinfo {author} {\bibfnamefont {M.}~\bibnamefont {Strickland}}, \bibinfo {author} {\bibfnamefont {S.}~\bibnamefont {Thapa}}, \bibinfo {author} {\bibfnamefont {P.}~\bibnamefont {Vander~Griend}}, \ and\ \bibinfo {author} {\bibfnamefont {J.~H.}\ \bibnamefont {Weber}},\ }\href {\doibase 10.1016/j.cpc.2021.108266} {\bibfield  {journal} {\bibinfo  {journal} {Comput. Phys. Commun.}\ }\textbf {\bibinfo {volume} {273}},\ \bibinfo {pages} {108266} (\bibinfo {year} {2022})},\ \Eprint {http://arxiv.org/abs/2107.06147} {arXiv:2107.06147} \BibitemShut {NoStop}%
\bibitem [{\citenamefont {Lindblad}(1976)}]{Lindblad:1975ef}%
  \BibitemOpen
  \bibfield  {author} {\bibinfo {author} {\bibfnamefont {G.}~\bibnamefont {Lindblad}},\ }\href {\doibase 10.1007/BF01608499} {\bibfield  {journal} {\bibinfo  {journal} {Commun. Math. Phys.}\ }\textbf {\bibinfo {volume} {48}},\ \bibinfo {pages} {119} (\bibinfo {year} {1976})}\BibitemShut {NoStop}%
\bibitem [{\citenamefont {Gorini}\ \emph {et~al.}(1976)\citenamefont {Gorini}, \citenamefont {Kossakowski},\ and\ \citenamefont {Sudarshan}}]{Gorini:1975nb}%
  \BibitemOpen
  \bibfield  {author} {\bibinfo {author} {\bibfnamefont {V.}~\bibnamefont {Gorini}}, \bibinfo {author} {\bibfnamefont {A.}~\bibnamefont {Kossakowski}}, \ and\ \bibinfo {author} {\bibfnamefont {E.~C.~G.}\ \bibnamefont {Sudarshan}},\ }\href {\doibase 10.1063/1.522979} {\bibfield  {journal} {\bibinfo  {journal} {J. Math. Phys.}\ }\textbf {\bibinfo {volume} {17}},\ \bibinfo {pages} {821} (\bibinfo {year} {1976})}\BibitemShut {NoStop}%
\bibitem [{\citenamefont {M\o{}lmer}\ \emph {et~al.}(1993)\citenamefont {M\o{}lmer}, \citenamefont {Castin},\ and\ \citenamefont {Dalibard}}]{Molmer:1993ltv}%
  \BibitemOpen
  \bibfield  {author} {\bibinfo {author} {\bibfnamefont {K.}~\bibnamefont {M\o{}lmer}}, \bibinfo {author} {\bibfnamefont {Y.}~\bibnamefont {Castin}}, \ and\ \bibinfo {author} {\bibfnamefont {J.}~\bibnamefont {Dalibard}},\ }\href {\doibase 10.1364/JOSAB.10.000524} {\bibfield  {journal} {\bibinfo  {journal} {J. Opt. Soc. Am. B}\ }\textbf {\bibinfo {volume} {10}},\ \bibinfo {pages} {524} (\bibinfo {year} {1993})}\BibitemShut {NoStop}%
\bibitem [{\citenamefont {Herren}\ and\ \citenamefont {Steinhauser}(2018)}]{Herren:2017osy}%
  \BibitemOpen
  \bibfield  {author} {\bibinfo {author} {\bibfnamefont {F.}~\bibnamefont {Herren}}\ and\ \bibinfo {author} {\bibfnamefont {M.}~\bibnamefont {Steinhauser}},\ }\href {\doibase 10.1016/j.cpc.2017.11.014} {\bibfield  {journal} {\bibinfo  {journal} {Comput. Phys. Commun.}\ }\textbf {\bibinfo {volume} {224}},\ \bibinfo {pages} {333} (\bibinfo {year} {2018})},\ \Eprint {http://arxiv.org/abs/1703.03751} {arXiv:1703.03751} \BibitemShut {NoStop}%
\bibitem [{\citenamefont {Anzai}\ \emph {et~al.}(2013)\citenamefont {Anzai}, \citenamefont {Prausa}, \citenamefont {Smirnov}, \citenamefont {Smirnov},\ and\ \citenamefont {Steinhauser}}]{Anzai:2013tja}%
  \BibitemOpen
  \bibfield  {author} {\bibinfo {author} {\bibfnamefont {C.}~\bibnamefont {Anzai}}, \bibinfo {author} {\bibfnamefont {M.}~\bibnamefont {Prausa}}, \bibinfo {author} {\bibfnamefont {A.~V.}\ \bibnamefont {Smirnov}}, \bibinfo {author} {\bibfnamefont {V.~A.}\ \bibnamefont {Smirnov}}, \ and\ \bibinfo {author} {\bibfnamefont {M.}~\bibnamefont {Steinhauser}},\ }\href {\doibase 10.1103/PhysRevD.88.054030} {\bibfield  {journal} {\bibinfo  {journal} {Phys. Rev. D}\ }\textbf {\bibinfo {volume} {88}},\ \bibinfo {pages} {054030} (\bibinfo {year} {2013})},\ \Eprint {http://arxiv.org/abs/1308.1202} {arXiv:1308.1202 [hep-ph]} \BibitemShut {NoStop}%
\bibitem [{\citenamefont {Pineda}(1998)}]{Pineda:1998PhD}%
  \BibitemOpen
  \bibfield  {author} {\bibinfo {author} {\bibfnamefont {A.}~\bibnamefont {Pineda}},\ }\emph {\bibinfo {title} {{Heavy Quarkonium and Nonrelativistic Effective Field Theories}}},\ \href {http://pineda.ifae.es/talks/main.ps} {Ph.D. thesis},\ \bibinfo  {school} {Unversitat de Barcelona} (\bibinfo {year} {1998})\BibitemShut {NoStop}%
\bibitem [{\citenamefont {Hoang}\ \emph {et~al.}(1999)\citenamefont {Hoang}, \citenamefont {Smith}, \citenamefont {Stelzer},\ and\ \citenamefont {Willenbrock}}]{Hoang:1998nz}%
  \BibitemOpen
  \bibfield  {author} {\bibinfo {author} {\bibfnamefont {A.~H.}\ \bibnamefont {Hoang}}, \bibinfo {author} {\bibfnamefont {M.~C.}\ \bibnamefont {Smith}}, \bibinfo {author} {\bibfnamefont {T.}~\bibnamefont {Stelzer}}, \ and\ \bibinfo {author} {\bibfnamefont {S.}~\bibnamefont {Willenbrock}},\ }\href {\doibase 10.1103/PhysRevD.59.114014} {\bibfield  {journal} {\bibinfo  {journal} {Phys. Rev. D}\ }\textbf {\bibinfo {volume} {59}},\ \bibinfo {pages} {114014} (\bibinfo {year} {1999})},\ \Eprint {http://arxiv.org/abs/hep-ph/9804227} {arXiv:hep-ph/9804227} \BibitemShut {NoStop}%
\bibitem [{\citenamefont {Bali}\ and\ \citenamefont {Pineda}(2004)}]{Bali:2003jq}%
  \BibitemOpen
  \bibfield  {author} {\bibinfo {author} {\bibfnamefont {G.~S.}\ \bibnamefont {Bali}}\ and\ \bibinfo {author} {\bibfnamefont {A.}~\bibnamefont {Pineda}},\ }\href {\doibase 10.1103/PhysRevD.69.094001} {\bibfield  {journal} {\bibinfo  {journal} {Phys. Rev. D}\ }\textbf {\bibinfo {volume} {69}},\ \bibinfo {pages} {094001} (\bibinfo {year} {2004})},\ \Eprint {http://arxiv.org/abs/hep-ph/0310130} {arXiv:hep-ph/0310130} \BibitemShut {NoStop}%
\bibitem [{\citenamefont {Zyla}\ \emph {et~al.}(2020)\citenamefont {Zyla} \emph {et~al.}}]{ParticleDataGroup:2020ssz}%
  \BibitemOpen
  \bibfield  {author} {\bibinfo {author} {\bibfnamefont {P.~A.}\ \bibnamefont {Zyla}} \emph {et~al.} (\bibinfo {collaboration} {Particle Data Group}),\ }\href {\doibase 10.1093/ptep/ptaa104} {\bibfield  {journal} {\bibinfo  {journal} {PTEP}\ }\textbf {\bibinfo {volume} {2020}},\ \bibinfo {pages} {083C01} (\bibinfo {year} {2020})}\BibitemShut {NoStop}%
\bibitem [{\citenamefont {Brambilla}\ \emph {et~al.}(2019)\citenamefont {Brambilla}, \citenamefont {Escobedo}, \citenamefont {Vairo},\ and\ \citenamefont {Vander~Griend}}]{Brambilla:2019tpt}%
  \BibitemOpen
  \bibfield  {author} {\bibinfo {author} {\bibfnamefont {N.}~\bibnamefont {Brambilla}}, \bibinfo {author} {\bibfnamefont {M.~A.}\ \bibnamefont {Escobedo}}, \bibinfo {author} {\bibfnamefont {A.}~\bibnamefont {Vairo}}, \ and\ \bibinfo {author} {\bibfnamefont {P.}~\bibnamefont {Vander~Griend}},\ }\href {\doibase 10.1103/PhysRevD.100.054025} {\bibfield  {journal} {\bibinfo  {journal} {Phys. Rev. D}\ }\textbf {\bibinfo {volume} {100}},\ \bibinfo {pages} {054025} (\bibinfo {year} {2019})},\ \Eprint {http://arxiv.org/abs/1903.08063} {arXiv:1903.08063 [hep-ph]} \BibitemShut {NoStop}%
\bibitem [{\citenamefont {Eller}\ \emph {et~al.}(2019)\citenamefont {Eller}, \citenamefont {Ghiglieri},\ and\ \citenamefont {Moore}}]{Eller:2019spw}%
  \BibitemOpen
  \bibfield  {author} {\bibinfo {author} {\bibfnamefont {A.~M.}\ \bibnamefont {Eller}}, \bibinfo {author} {\bibfnamefont {J.}~\bibnamefont {Ghiglieri}}, \ and\ \bibinfo {author} {\bibfnamefont {G.~D.}\ \bibnamefont {Moore}},\ }\href {\doibase 10.1103/PhysRevD.99.094042} {\bibfield  {journal} {\bibinfo  {journal} {Phys. Rev. D}\ }\textbf {\bibinfo {volume} {99}},\ \bibinfo {pages} {094042} (\bibinfo {year} {2019})},\ \bibinfo {note} {[Erratum: Phys.Rev.D 102, 039901 (2020)]},\ \Eprint {http://arxiv.org/abs/1903.08064} {arXiv:1903.08064 [hep-ph]} \BibitemShut {NoStop}%
\bibitem [{\citenamefont {Brambilla}\ \emph {et~al.}(2020)\citenamefont {Brambilla}, \citenamefont {Leino}, \citenamefont {Petreczky},\ and\ \citenamefont {Vairo}}]{Brambilla:2020siz}%
  \BibitemOpen
  \bibfield  {author} {\bibinfo {author} {\bibfnamefont {N.}~\bibnamefont {Brambilla}}, \bibinfo {author} {\bibfnamefont {V.}~\bibnamefont {Leino}}, \bibinfo {author} {\bibfnamefont {P.}~\bibnamefont {Petreczky}}, \ and\ \bibinfo {author} {\bibfnamefont {A.}~\bibnamefont {Vairo}},\ }\href {\doibase 10.1103/PhysRevD.102.074503} {\bibfield  {journal} {\bibinfo  {journal} {Phys. ~Rev. ~D}\ }\textbf {\bibinfo {volume} {102}},\ \bibinfo {pages} {074503} (\bibinfo {year} {2020})},\ \Eprint {http://arxiv.org/abs/2007.10078} {arXiv:2007.10078 [hep-lat]} \BibitemShut {NoStop}%
\bibitem [{\citenamefont {Foster}\ and\ \citenamefont {Michael}(1999)}]{Foster:1998wu}%
  \BibitemOpen
  \bibfield  {author} {\bibinfo {author} {\bibfnamefont {M.}~\bibnamefont {Foster}}\ and\ \bibinfo {author} {\bibfnamefont {C.}~\bibnamefont {Michael}} (\bibinfo {collaboration} {UKQCD}),\ }\href {\doibase 10.1103/PhysRevD.59.094509} {\bibfield  {journal} {\bibinfo  {journal} {Phys. Rev. D}\ }\textbf {\bibinfo {volume} {59}},\ \bibinfo {pages} {094509} (\bibinfo {year} {1999})},\ \Eprint {http://arxiv.org/abs/hep-lat/9811010} {arXiv:hep-lat/9811010} \BibitemShut {NoStop}%
\bibitem [{\citenamefont {Narison}(2023)}]{Narison:2023npv}%
  \BibitemOpen
  \bibfield  {author} {\bibinfo {author} {\bibfnamefont {S.}~\bibnamefont {Narison}},\ }in\ \href@noop {} {\emph {\bibinfo {booktitle} {{26th High-Energy Physics International Conference in QCD}}}}\ (\bibinfo {year} {2023})\ \Eprint {http://arxiv.org/abs/2309.05342} {arXiv:2309.05342 [hep-ph]} \BibitemShut {NoStop}%
\bibitem [{\citenamefont {Herr}\ \emph {et~al.}(2024)\citenamefont {Herr}, \citenamefont {Schlosser},\ and\ \citenamefont {Wagner}}]{Herr:2023xwg}%
  \BibitemOpen
  \bibfield  {author} {\bibinfo {author} {\bibfnamefont {J.}~\bibnamefont {Herr}}, \bibinfo {author} {\bibfnamefont {C.}~\bibnamefont {Schlosser}}, \ and\ \bibinfo {author} {\bibfnamefont {M.}~\bibnamefont {Wagner}},\ }\href {\doibase 10.1103/PhysRevD.109.034516} {\bibfield  {journal} {\bibinfo  {journal} {Phys. Rev. D}\ }\textbf {\bibinfo {volume} {109}},\ \bibinfo {pages} {034516} (\bibinfo {year} {2024})},\ \Eprint {http://arxiv.org/abs/2306.09902} {arXiv:2306.09902 [hep-lat]} \BibitemShut {NoStop}%
\bibitem [{\citenamefont {Alqahtani}\ and\ \citenamefont {Strickland}(2021)}]{Alqahtani:2020paa}%
  \BibitemOpen
  \bibfield  {author} {\bibinfo {author} {\bibfnamefont {M.}~\bibnamefont {Alqahtani}}\ and\ \bibinfo {author} {\bibfnamefont {M.}~\bibnamefont {Strickland}},\ }\href {\doibase 10.1140/epjc/s10052-021-09832-z} {\bibfield  {journal} {\bibinfo  {journal} {Eur. Phys. J. C}\ }\textbf {\bibinfo {volume} {81}},\ \bibinfo {pages} {1022} (\bibinfo {year} {2021})},\ \Eprint {http://arxiv.org/abs/2008.07657} {arXiv:2008.07657 [nucl-th]} \BibitemShut {NoStop}%
\bibitem [{\citenamefont {Brambilla}\ \emph {et~al.}(2021{\natexlab{b}})\citenamefont {Brambilla}, \citenamefont {Escobedo}, \citenamefont {Strickland}, \citenamefont {Vairo}, \citenamefont {Vander~Griend},\ and\ \citenamefont {Weber}}]{Brambilla:2021wkt}%
  \BibitemOpen
  \bibfield  {author} {\bibinfo {author} {\bibfnamefont {N.}~\bibnamefont {Brambilla}}, \bibinfo {author} {\bibfnamefont {M.~A.}\ \bibnamefont {Escobedo}}, \bibinfo {author} {\bibfnamefont {M.}~\bibnamefont {Strickland}}, \bibinfo {author} {\bibfnamefont {A.}~\bibnamefont {Vairo}}, \bibinfo {author} {\bibfnamefont {P.}~\bibnamefont {Vander~Griend}}, \ and\ \bibinfo {author} {\bibfnamefont {J.~H.}\ \bibnamefont {Weber}},\ }\href {\doibase 10.1103/PhysRevD.104.094049} {\bibfield  {journal} {\bibinfo  {journal} {Phys. Rev. D}\ }\textbf {\bibinfo {volume} {104}},\ \bibinfo {pages} {094049} (\bibinfo {year} {2021}{\natexlab{b}})},\ \Eprint {http://arxiv.org/abs/2107.06222} {arXiv:2107.06222} \BibitemShut {NoStop}%
\bibitem [{\citenamefont {Brambilla}\ \emph {et~al.}(2002)\citenamefont {Brambilla}, \citenamefont {Sumino},\ and\ \citenamefont {Vairo}}]{Brambilla:2001qk}%
  \BibitemOpen
  \bibfield  {author} {\bibinfo {author} {\bibfnamefont {N.}~\bibnamefont {Brambilla}}, \bibinfo {author} {\bibfnamefont {Y.}~\bibnamefont {Sumino}}, \ and\ \bibinfo {author} {\bibfnamefont {A.}~\bibnamefont {Vairo}},\ }\href {\doibase 10.1103/PhysRevD.65.034001} {\bibfield  {journal} {\bibinfo  {journal} {Phys. Rev. D}\ }\textbf {\bibinfo {volume} {65}},\ \bibinfo {pages} {034001} (\bibinfo {year} {2002})},\ \Eprint {http://arxiv.org/abs/hep-ph/0108084} {arXiv:hep-ph/0108084} \BibitemShut {NoStop}%
\bibitem [{\citenamefont {Fischler}(1977)}]{Fischler:1977yf}%
  \BibitemOpen
  \bibfield  {author} {\bibinfo {author} {\bibfnamefont {W.}~\bibnamefont {Fischler}},\ }\href {\doibase 10.1016/0550-3213(77)90026-8} {\bibfield  {journal} {\bibinfo  {journal} {Nucl. Phys. B}\ }\textbf {\bibinfo {volume} {129}},\ \bibinfo {pages} {157} (\bibinfo {year} {1977})}\BibitemShut {NoStop}%
\bibitem [{\citenamefont {Billoire}(1980)}]{Billoire:1979ih}%
  \BibitemOpen
  \bibfield  {author} {\bibinfo {author} {\bibfnamefont {A.}~\bibnamefont {Billoire}},\ }\href {\doibase 10.1016/0370-2693(80)90279-8} {\bibfield  {journal} {\bibinfo  {journal} {Phys. Lett. B}\ }\textbf {\bibinfo {volume} {92}},\ \bibinfo {pages} {343} (\bibinfo {year} {1980})}\BibitemShut {NoStop}%
\bibitem [{\citenamefont {Peter}(1997)}]{Peter:1997me}%
  \BibitemOpen
  \bibfield  {author} {\bibinfo {author} {\bibfnamefont {M.}~\bibnamefont {Peter}},\ }\href {\doibase 10.1016/S0550-3213(97)00373-8} {\bibfield  {journal} {\bibinfo  {journal} {Nucl. Phys. B}\ }\textbf {\bibinfo {volume} {501}},\ \bibinfo {pages} {471} (\bibinfo {year} {1997})},\ \Eprint {http://arxiv.org/abs/hep-ph/9702245} {arXiv:hep-ph/9702245} \BibitemShut {NoStop}%
\bibitem [{\citenamefont {Schroder}(1999)}]{Schroder:1998vy}%
  \BibitemOpen
  \bibfield  {author} {\bibinfo {author} {\bibfnamefont {Y.}~\bibnamefont {Schroder}},\ }\href {\doibase 10.1016/S0370-2693(99)00010-6} {\bibfield  {journal} {\bibinfo  {journal} {Phys. Lett. B}\ }\textbf {\bibinfo {volume} {447}},\ \bibinfo {pages} {321} (\bibinfo {year} {1999})},\ \Eprint {http://arxiv.org/abs/hep-ph/9812205} {arXiv:hep-ph/9812205} \BibitemShut {NoStop}%
\bibitem [{\citenamefont {Ayala}\ \emph {et~al.}(2014)\citenamefont {Ayala}, \citenamefont {Cveti\v{c}},\ and\ \citenamefont {Pineda}}]{Ayala:2014yxa}%
  \BibitemOpen
  \bibfield  {author} {\bibinfo {author} {\bibfnamefont {C.}~\bibnamefont {Ayala}}, \bibinfo {author} {\bibfnamefont {G.}~\bibnamefont {Cveti\v{c}}}, \ and\ \bibinfo {author} {\bibfnamefont {A.}~\bibnamefont {Pineda}},\ }\href {\doibase 10.1007/JHEP09(2014)045} {\bibfield  {journal} {\bibinfo  {journal} {JHEP}\ }\textbf {\bibinfo {volume} {09}},\ \bibinfo {pages} {045} (\bibinfo {year} {2014})},\ \Eprint {http://arxiv.org/abs/1407.2128} {arXiv:1407.2128 [hep-ph]} \BibitemShut {NoStop}%
\bibitem [{\citenamefont {Komijani}(2017)}]{Komijani:2017vep}%
  \BibitemOpen
  \bibfield  {author} {\bibinfo {author} {\bibfnamefont {J.}~\bibnamefont {Komijani}},\ }\href {\doibase 10.1007/JHEP08(2017)062} {\bibfield  {journal} {\bibinfo  {journal} {JHEP}\ }\textbf {\bibinfo {volume} {08}},\ \bibinfo {pages} {062} (\bibinfo {year} {2017})},\ \Eprint {http://arxiv.org/abs/1701.00347} {arXiv:1701.00347 [hep-ph]} \BibitemShut {NoStop}%
\end{thebibliography}%

\end{document}